\documentclass[aps,prx,reprint,preprintnumbers,superscriptaddress,nofootinbib,longbibliography,floatfix]{revtex4-2}
\pdfoutput=1
\usepackage{rotating}
\usepackage{array}
\usepackage{amsmath}
\usepackage[normalem]{ulem}
\usepackage{slashed}
\usepackage{booktabs}
\usepackage[pdftex,table]{xcolor}
\usepackage{units}
\usepackage{xfrac}
\usepackage{mathtools}
\usepackage{empheq}
\usepackage[]{units}
\usepackage{multirow}
\usepackage{amssymb}
\usepackage{url}
\usepackage{comment}
\usepackage{physics}
\usepackage{color,soul}
\usepackage{bbm}
\usepackage[caption=false]{subfig}
\usepackage{adjustbox}
\usepackage[T1]{fontenc}

\usepackage{hyperref}
\hypersetup{
  colorlinks=true,
  citecolor=blue,
  linkcolor=blue,
  urlcolor=blue
}

\newcommand{\pt}{\mathrm{p_{T}}}

\begin{document}

\title{A Method to Simultaneously Facilitate All Jet Physics Tasks}

\author{Vinicius Mikuni}
\email{vmikuni@lbl.gov}
\affiliation{National Energy Research Scientific Computing Center, Berkeley Lab, Berkeley, CA 94720, USA}

\author{Benjamin Nachman}
\email{bpnachman@lbl.gov}
\affiliation{Physics Division, Lawrence Berkeley National Laboratory, Berkeley, CA 94720, USA}
\affiliation{Berkeley Institute for Data Science, University of California, Berkeley, CA 94720, USA}

\begin{abstract}
    Machine learning has become an essential tool in jet physics.  Due to their complex, high-dimensional nature, jets can be explored holistically by neural networks in ways that are not possible manually.  However, innovations in all areas of jet physics are proceeding in parallel.  We show that specially constructed machine learning models trained for a specific jet classification task can improve the accuracy, precision, or speed of all other jet physics tasks.  This is demonstrated by training on a particular multiclass generation and classification task and then using the learned representation for different generation and classification tasks, for datasets with a different (full) detector simulation, for jets from a different collision system ($pp$ versus $ep$), for generative models, for likelihood ratio estimation, and for anomaly detection. We consider, our \textsc{OmniLearn} approach thus as a jet-physics foundation model. It is made publicly available for use in any area where state-of-the-art precision is required for analyses involving jets and their substructure.
\end{abstract}

\maketitle

{\small
\tableofcontents
}

\vspace{10mm}

\section{Introduction}
\label{sec:intro}

The study of high-energy hadronic final states -- jet physics -- is seeing a paradigm shift from modern machine learning (ML).  Jets are composed of many particles, each with properties of their own.  This means that jets are represented in high-dimensional spaces and are thus difficult to analyze manually.  There has been incredible progress over the last two decades in classical jet physics to develop observables and other techniques using direct physics reasoning~\cite{Abdesselam:2010pt,Altheimer:2012mn,Altheimer:2013yza,Adams:2015hiv,Larkoski:2017jix,Kogler:2018hem,Marzani:2019hun,Kogler:2021kkw}, but the deep learning revolution of the last few years has shown that automation and indirect physics reasoning (e.g. through simulations and general physics considerations) can significantly improve performance in many tasks.  For example, classifying a jet as originating from a top quark or a generic quark/gluon jet is over 20 times more effective with the latest deep learning solutions compared with classical methods~\cite{Kasieczka:2019dbj,Gong:2022lye,Bogatskiy:2023nnw}.  Deep learning has also enabled new studies that were unimaginable before, like unbinned differential cross sections in tens or hundreds of dimensions simultaneously~\cite{Andreassen:2019cjw,Bellagente:2019uyp,1800956,Vandegar:2020yvw,Andreassen:2021zzk,Arratia:2021otl,Howard:2021pos,Backes:2022vmn,Shmakov:2023kjj,Shmakov:2024gkd}. 
However, one feature these innovations have in common is that they are all advancing in parallel.  Our question is simple: \textit{can we make progress on all jet physics tasks at the same time?}

One answer to this question has been inspired by recent progress in large language modeling (LLMs).  Tools like ChatGPT (\url{https://chat.openai.com}) and others~\cite{zhou2023comprehensive} are called \textit{foundation models} because they are able to approach many downstream tasks either with little or no finetuning to the specific problem.  They have many millions (or billions/trillions) parameters and are trained using many millions (or much more) of examples.  
Recently, LLMs have also been adapted for particle physics-specific queries~\cite{Zhang:2024kws}.  Such tools complement our approach: they target the process of particle physics research while we focus on processing particle physics data.

Foundation models are usually trained using a form of self-supervision (e.g. mask data and learn to fill in the blanks) in order to learn to represent the structure of data.  This sort of representation learning has been studied recently in particle physics using a number of interesting approaches~\cite{Dillon:2021gag,Dillon:2022tmm,Dillon:2023zac,Heinrich:2024sbg,Harris:2024sra,Kuh:2024lgx}.  As desired, the learned representations are also demonstrated to improve downstream tasks. However, the way these models are trained is not aligned with any actual analysis goal (we do not actually want to fill in blanks).  Our intuition is that foundation models are useful because they increase the effective size of the training dataset for a downstream task.  The closer the foundation model training is to the downstream task, the larger the increase in the effective size for a fixed sample size used to train the foundation model.  A key difference between foundation models in (particle) physics and foundation models in society at large is the existence of \textit{ab initio} simulations.  These simulations provide large datasets that can be used to target specific tasks.  Since we are in a privileged situation in which machine learning models can be constructed for dedicated problems, \textit{can such a ML model dedicated to a specific jet physics task act as a foundation model?}  

The process of (pre)training an ML model for one task and then applying it elsewhere, usually with some fine-tuning to the downstream task, is called \textit{transfer learning}.  This strategy has been shown to be effective in a number of applications across particle physics~\cite{Kuchera:2018djs,Chappell:2022yxd,Dreyer:2022yom,Beauchesne:2023vie}.  In all of these cases, there was only one downstream task.  Furthermore, either the pretraining was very far from the target task (e.g. pretraining with generic non-physics images) or very close to the target task (e.g. classify jet of type A and transfer to type B).  These papers were not trying to build foundation models, but we hypothesize that a significantly scaled up version of the transfer learning task can form the basis for a foundation model in particle physics. As such, our goal is to see if we can build a foundation model for jet physics by using a supervised instead of self-supervised learning.  Instead of predicting tokens, we develop a new pre-training strategy that leverages the simulation information available from collider physics and our ability to generate labeled data. We pair that strategy with an unsupervised strategy that aims to generate jets using a diffusion process. Contrary to token prediction, generative models are often used in applications such as anomaly detection and hence better aligned with our physics goals. Our neural network will have many millions of parameters and will be trained on 100 million jets.  We will achieve success if the implicit representation learned by our model improves (in training speed, accuracy, or both) essentially all downstream tasks in jet physics.  

An approach with a similar philosophy to ours is \textsc{OmniJet}-$\alpha$~\cite{Birk:2024knn}.  The authors of \textsc{OmniJet}-$\alpha$ use a generative model trained on a specific jet physics task in order to build a foundation model.  A key difference is that \textsc{OmniJet}-$\alpha$ is based on a language model and so jet constituents are discretized (into `tokens') and the generation is autoregressive (like generating a sentence from left to right).  On the application side, \textsc{OmniJet}-$\alpha$ was trained on one generative task and applied on one classification task using the same dataset, whereas we try to explore most areas of jet physics across a variety of datasets with our model. 

Our paper is organized as follows.  Section~\ref{sec:pet} describes how jets can be represented as point clouds and the neural network architecture at the center of our foundation model. Just how \textsc{OmniFold}~\cite{Andreassen:2019cjw} can unfold all dimensions simultaneously, our approach -- called \textsc{OmniLearn} -- can learn useful representations for all jet physics tasks simultaneously.  This is demonstrated in subsequent sections by showing that our model can enhance and/or accelerate tasks other than the one it was trained on (multi-class jets from fast simulation) to binary classification on a different dataset (Sec.~\ref{sec:top_qg}), to binary classification using a full detector simulation (Sec.~\ref{sec:lhc}), to jets originating from a different collision system (Sec.~\ref{sec:h1}), to generative models (Sec.~\ref{sec:jetnet}), to likelihood ratio estimation (Sec.~\ref{sec:omnifold}), and to anomaly detection (Sec.~\ref{sec:lhco}).  The paper ends with conclusions and outlook in Sec.~\ref{sec:conclusions}.

\section{Point Clouds for Jet Physics}
\label{sec:pet}

\begin{figure*}[ht]
    \centering
        \includegraphics[width=.95\textwidth]{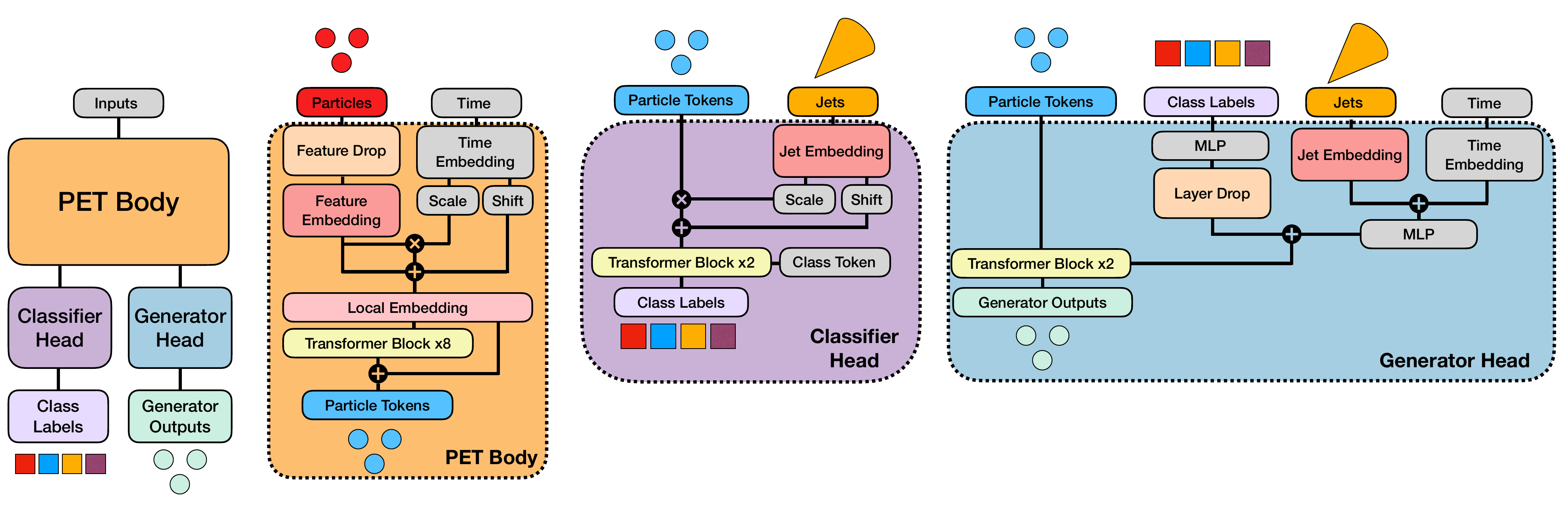}
    \caption{Neural network architecture used to train \textsc{OmniLearn}. The main neural network blocks of the architecture are shown in the further left with detailed architecture design shown for each block in the right. See the text for more details.}
    \label{fig:pet}
\end{figure*}

The interpretation of jets as point clouds motivated the development of new neural network architectures to naturally address the data structure represented by an unordered set of particles with varying number of constituents. Successful neural network models using sets~\cite{Komiske:2018cqr,ATL-PHYS-PUB-2020-014}, graph neural networks (GNNs)~\cite{Moreno:2019bmu,Qu:2019gqs,Shimmin:2021pkm}, and more recently transformers~\cite{Mikuni:2020wpr,Mikuni:2021pou,Qu:2022mxj}, were all able to successfully improve upon previous deep learning approaches in collider physics. The first transformer model in high energy physics~\cite{Mikuni:2020wpr} used graph-attention networks~\cite{velivckovic2017graph}, combining the advantages of both GNNs and transformers. The rise in popularity of transformer models is also attributed to their strong scaling properties~\cite{DBLP:journals/corr/abs-2001-08361}, making it the backbone of almost all modern foundational models and the choice of neural network architecture used to build \textsc{OmniLearn}.

For collider physics applications, we expect that in the presence of large datasets and suitable tasks, a foundational model for jets will be able to leverage the information to learn a general representation of jets, thus providing a stronger inductive bias that will be transferable to different datasets and tasks. We also notice that the overwhelming majority of machine learning applications for jet physics can be summarized as a classification or generation tasks. These include the applications in jet tagging and jet generation as well as complex tasks such as event reweighting and anomaly detection. Motivated by this observation, we conjecture that a flexible model, capable of learning both to generate and classify jets, will also learn a useful and general data representation that can be employed to quickly adapt to new downstream datasets and applications. 

While multiple generative models have been proposed for point cloud generation in particle physics based on generative adversarial networks~\cite{Kansal:2021cqp,Buhmann:2023pmh}, variational autoencoders~\cite{Touranakou:2022qrp}, normalizing flows~\cite{Kach:2022qnf,Verheyen:2022tov}, and auto-regressive transformers~\cite{Finke:2023veq,Butter:2023fov},  we choose to focus on diffusion generative models~\cite{sohldickstein2015deep}. Diffusion generative models use time-dependent perturbations applied to the data to learn an approximation of the score function, or gradients of the logarithm of the probability density of the data. This choice also aligns with the classification task. Since the perturbation process is designed such that no perturbation is applied at the initial time $t=0$, conditioning the model over the time parameter ensures the network is able to accommodate both perturbed and unperturbed data simultaneously. To build \textsc{OmniLearn}, we design the architecture with a shared representation whose outputs are then fed to task-specific neural networks. This approach enables flexibility and efficient design, since downstream task applications only need to load the shared representation and relevant task-specific network, reducing the overall model. The main building blocks of the network are summarized in Fig.~\ref{fig:pet}. In the following subsections we will provide a detailed description of the model and the core design choices.

\subsection{Point-Edge Transformer}
\label{sec:petbody}

The shared representation of the network takes as inputs the particles clustered inside the jets and is conditioned on the diffusion time parameter. The time information, following previous diffusion models for collider physics~\cite{Mikuni:2023dvk,Buhmann:2023acn,Mikuni:2023tok,Leigh:2023toe,Shmakov:2023kjj}, is encoded to a higher dimensional space using a time embedding layer. The time embedding consists Fourier features \cite{tancik2020fourier} followed by two multi-layer perceptrons (MLPs) with \textsc{GELU} activation function~\cite{hendrycks2016gaussian}. Unless otherwise stated, all MLP layers used in this work are followed by a \textsc{GELU} non-linear activation. Contrary to previous diffusion models, we modify the time embedding by multiplying the output of the Fourier features by the time parameter, such that the output of the time embedding is zero when the input time is also zero. This choice ensures the time embedding is effectively turned off when the model is evaluated in classifier mode. The next step is to combine the time information with the input particle information. Datasets store different levels of information for each particle. The most basic information, described by the kinematic information of each particle, is always stored. However, additional information such as particle identification (PID) and vertex information for charged particles is only available in specific datasets, with the latter only provided in the JetClass dataset~\cite{Qu:2022mxj} among benchmark, public jet datasets. To avoid training multiple models to accommodate each dataset, and thus defeating the purpose of a generalized model, we instead adopt a feature drop approach. During training, we consider as inputs both the kinematic information for each particle and their respective PID\footnote{The vertex information while present in the JetClass dataset is not used during training. While that could also be included in the training methodology, our focus is the application to multiple datasets without this information, hence for simplicity we skip these features. }. With a probability $p = 0.2$ we drop the PID information by replacing it with zeros. This approach is similar to dropout layers~\cite{JMLR:v15:srivastava14a} that encourage the network to learn a useful representation both in the presence and absence of these features. After the feature drop, the inputs are encoded to a higher dimensional space using a feature embedding consisting of two MLP layers. The outputs of the feature embedding are then combined with the time information though a shift and scaling operation. Before the transformer block we introduce a positional token to encode the geometrical information of the neighborhood surrounding each particle inside the jet. Even though transformers are capable of learning general correlations between particles, the addition of local information can generally improve performance~\cite{Mikuni:2021pou}, creating a better latent representation that is aware of the distances between particles. We create the local encoding using dynamic graph convolution (DGCNNs)~\cite{DBLP:journals/corr/abs-1801-07829} layers where the neighborhood is defined using a $k$-nearest neighbor algorithm with number of neighbors fixed to 10. The distances are calculated in the pseudorapidity-azimuthal angle space. For each of the $k$-neighbors, edge features are defined based on the particle features concatenated with the subtraction between the particle features and each of the respective neighbors. An MLP is used over all edges before an average pooling operation over the neighbor dimension. The result of the operation represents the new particle features. A second DGCNN layer is then created  with distances calculated based on the Euclidean distances between particles after the updated features. This second operation allows the dynamic construction of edges between particles in a learned geometrical space that pulls relevant particles closer together and far apart otherwise.  The particle embedding equipped with local information is then passed through multiple transformer blocks. The transformer block closely follows the original proposal~\cite{DBLP:journals/corr/VaswaniSPUJGKP17}, combining multi-head attention modules with additional skip connections. Our main modification is the addition of \textsc{LayerScale}~\cite{touvron2021going} layers. The \textsc{LayerScale} operation introduces a learnable multiplicative factor to each skip connection in the transformer block. Using a small initial value ($10^{-5}$ in our implementation), the operation improves the stability and convergence of the model by allowing each transformer block to learn the relevance of each attention block in the transformer architecture. The outputs of the transformer blocks are then added to the original particle embedding after the combination with the local tokens, improving the information flow over the entire model. The outputs are then used as inputs of the task-specific blocks. Since our proposed backbone architecture combines edge creation with transformer modules we refer to it as Point-Edge Transformer (PET), noting that \textsc{OmniLearn} refers to the join training strategy, while PET refers to the specific neural network architecture used to implement \textsc{OmniLearn}.

\subsection{Classifier Head}
The classification task across different jet types is accomplished through a dedicated classification head that takes as inputs the particles after the shared network, now referred to as the \textsc{PET} body. Additionally, we also include the overall information from the jet kinematics, including the jet mass, transverse momentum $\pt$, pseudorapidity $\eta$, and particle multiplicity. Even though the additional information is partially redundant compared to the initial particle features, the addition of the jet kinematic information helps the model converge faster when evaluated over datasets covering different fiducial regions than the ones used during training. The jet information is embedded in a higher dimensional space using a jet embedding layer that includes two MLP layers of same size as the current particle embedding dimensionality. This information is then combined with the outputs of the \textsc{PET} body through a scaling and shift operations. A trainable class token~\cite{dosovitskiy2020image} is then used to summarize the information of the particle embeddings before the classification output. The class token is essentially interpreted as an additional particle, concatenated to the true particle inputs. Inside the transformer block, the outputs of the \textsc{PET} body are not updated but only the class token is allowed to change at the end of each transformer block. The output predictions are then determined by passing the updated class tokens over one last MLP with output size determined by the number of classes in the dataset.

\subsection{Generator Head}
Similarly to the classification head, the generator head takes as inputs both particle embeddings and jet kinematic information. Additionally, we include the time information and the set of class labels to condition the generator over the jet types to be simulated. The time and jet information are embedded in a higher dimensional space using the same encoding blocks used in the \textsc{PET} body and classifier head, respectively. This information is then combined through an addition operation followed by an MLP. The classification labels are also mapped to a higher dimensional space using a single MLP. The outputs of the MLP are then passed through a layer drop operation that, similarly to the feature drop, has a probability $p = 0.1$ to replace the entire output of the MLP with zeros. This choice is motivated by two observations: when the model is used during downstream tasks, the classes used to condition the \textsc{PET} architecture are hardly going to be the same as the ones used during training. Randomly ignoring the class labels encourages the entire architecture to learn both a general and specialized representation, leading to quicker convergence when adapted to other datasets. Second, this technique is similar to classifier-free guidance, observed to improve the generation quality of diffusion models~\cite{ho2022classifier}.
The results of the layer drop operation are then added to the  outputs of the combined jet and time embeddings. The result of this combination is then used as a diffusion token, where similarly to the classification token, is tasked to summarize the particle embedding information inside the transformer block. However, while the classification token is interpreted as an additional particle, we interpret the diffusion token as a conditional shift of the particle embeddings produced by the \textsc{PET} body. Initially, all particles are simultaneously shifted by the diffusion token created from the combined class labels, time, and jet information. The diffusion tokens are then updated with every transformer layer. The diffusion prediction is then the sum of the original \textsc{PET} body outputs with the learned diffusion tokens. 

\subsection{Loss Function}
\label{sec:loss}
The loss function of \textsc{OmniLearn} consists of multiple terms designed to combine both the classification and generation tasks. Each of the terms is shown in Equation~\ref{eq:loss}.

\begin{equation}
\begin{aligned}
    \mathcal{L} &= \mathcal{L}_{\text{class}} + \mathcal{L}_{\text{gen}} + \mathcal{L}_{\text{class smear}} \\
                &= \textsc{CE}(y,y_{\text{pred}}) +  \left\| \mathbf{v} - \mathbf{v}_{\text{pred}}\right\|^2 + \alpha^2\textsc{CE}(y,\hat{y}_{\text{pred}})\,.
\end{aligned}
\label{eq:loss}
\end{equation}

For input data $\textbf{x}$, the cross-entropy (CE) loss is calculated using the output of the \textsc{PET} classifier $y_{\text{pred}}$ and true class labels $y$. The \textsc{PET} generator takes as inputs perturbed data $\hat{\textbf{x}} = \alpha(t)\textbf{x} + \sigma(t)\boldsymbol{\epsilon}$ with time-dependent perturbation parameters $\alpha(t)$ and $\sigma(t)$ and predicts a velocity parameter $\textbf{v}_{\text{pred}}$ that is compared with the true velocity value $\textbf{v} = \alpha(t)\boldsymbol{\epsilon} - \sigma(t)\textbf{x}$. Additionally, the perturbed inputs can also be interpreted as a form of data augmentation that can further improve the classifier performance. We evaluate the \textsc{PET} classifier over the perturbed inputs to get the predictions $\hat{y}_{\text{pred}}$ used in the calculation of the cross entropy loss. A weight of $\alpha(t)^2$ is applied to ensure that at $t=0$, where $\alpha(0) = 1$ and $\sigma(0) = 0$, we recover the classifier loss over clean inputs and at $t=1$, where $\alpha(1) = 0$ and $\sigma(1) = 1$, the completely corrupted data does not negatively impact the classification performance.

\subsection{Training Details}
\label{sec:training}

We train \textsc{OmniLearn} using the JetClass dataset~\cite{Qu:2022mxj}. A total of 10 different jet classes are provided, simulated using MADGRAPH5\_aMC@NLO~\cite{Alwall:2014hca} for the matrix element calculation and \textsc{Pythia8}~\cite{Sjostrand:2006za,Sjostrand:2014zea} to perform the parton showering and hadronization. Detector effects are simulated using \textsc{Delphes3.4.3}~\cite{deFavereau:2013fsa,Mertens:2015kba,Selvaggi:2014mya} with the CMS detector configuration. Jets are clustered using the anti-$k_T$ algorithm with radius parameter of $R=0.8$.~\cite{Cacciari:2005hq,Cacciari:2011ma,Cacciari:2008gp}. Jets with transverse momentum between 500-1000 GeV and pseudorapidity $|\eta| < 2.0$ are saved. The training dataset consists of 100M jets, divided equally between each of the 10 classes. The kinematic information and PID for each particle is used as input features for the training and listed in Appendix~\ref{app:inputs}.

Up to 150 particles are saved per jet to be used during training. The training is carried out on the Perlmutter Supercomputer~\cite{Perlmutter} using 128 GPUs simultaneously with Horovod~\cite{sergeev2018horovod} package for data distributed training. A local batch size of size 256 is used with model training up to 200 epochs. \textsc{OmniLearn} is implemented in \textsc{TensorFlow}~\cite{tensorflow} with \textsc{Keras}~\cite{keras} backend. The cosine learning rate schedule~\cite{DBLP:journals/corr/LoshchilovH16a} is used with an initial learning rate of $3\times10^{-5}$, increasing to $3\sqrt{128}\times10^{-5}$ after three epochs and decreasing to $10^{-6}$ until the end of the training. The \textsc{Lion} optimizer~\cite{chen2024symbolic} is used with parameters $\beta_1 = 0.95$ and $\beta_2 = 0.99$. The fine-tuning of \textsc{OmniLearn} across different datasets and tasks is performed by setting the learning rate of the \textsc{PET} body to be a factor 10 smaller than the learning rate used in the rest of the \textsc{PET} architecture. The \textsc{PET} body model has 1.3M trainable weights, while the classifier and generator heads have 268k and 416k trainable parameters, respectively. 

In every application, all pre-trained weights are loaded unless there is a change in input/output dimensions. This is only the case for the output layer of the classification model and the input class labels in the generator model, where the number of classes used in each downstream task changes. In these cases, a new layer with correct input and output sizes replaces the trained weights and is initialized using random weights. Notice that the model itself does not rely on the specific particle multiplicity of the input point cloud and no changes to the architecture are necessary to address the different multiplicities between datasets. We evaluate \textsc{OmniLearn} across 9 different datasets with results described in the following sections.

\section{Generalization across Jet Types}
\label{sec:top_qg}

\begin{figure*}[th]
    \centering
        \includegraphics[width=.4\textwidth]{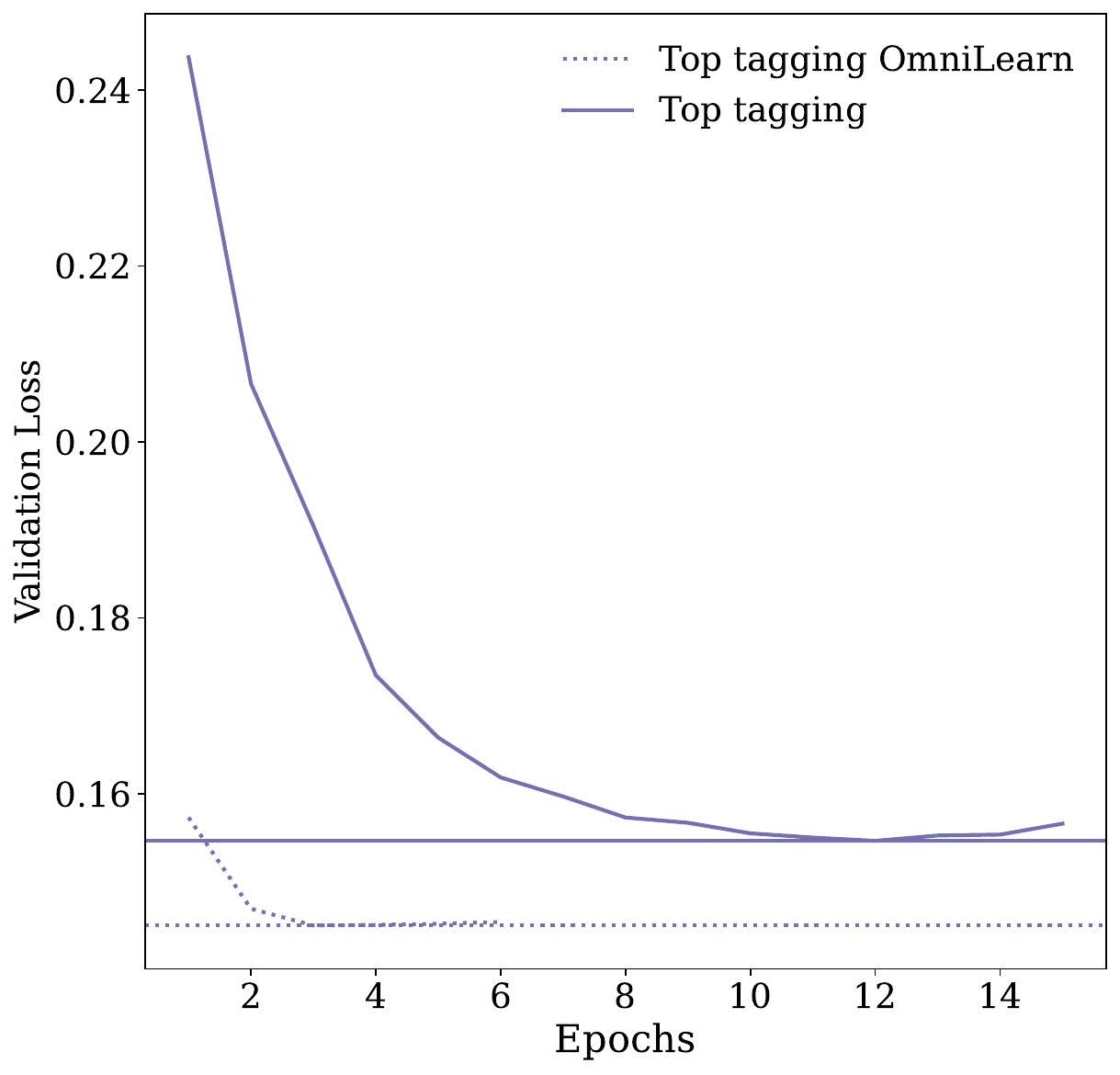}
        \includegraphics[width=.4\textwidth]{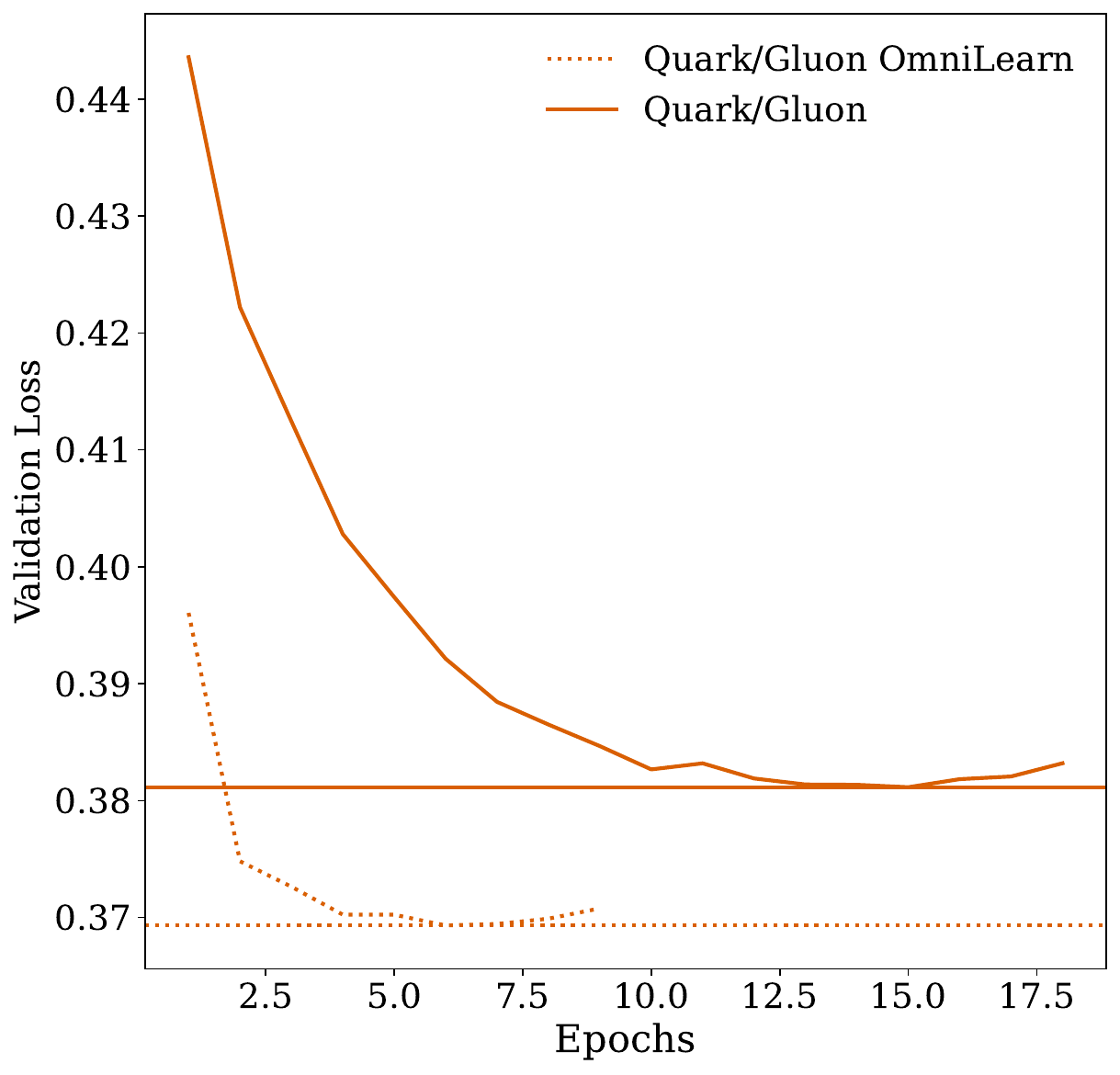}
    \caption{Validation loss curves obtained in the top quark tagging (left) and quark/ gluon (rights) datasets. The \textsc{OmniLearn} validation loss is compared with the \textsc{PET} classifier trained from scratch.}
    \label{fig:loss_top_qg}
\end{figure*}

\begin{table}[th]
    \centering
    \caption{Comparison between the performance reported for different classification algorithms on the top tagging dataset. The uncertainty quoted corresponds to the standard deviation of five trainings with different random weight initialization. If the uncertainty is not quoted then the variation is negligible compared to the expected value. Bold results represent the algorithm with highest performance. }
    \label{tab:results_top}
	\begin{tabular}{lccccc}
    \noalign{\smallskip}\hline
          &  Acc &AUC & \multicolumn{2}{c}{1/$\epsilon_B$} \\
          \cline{4-5}
          & & & $\epsilon_S = 0.5$ & $\epsilon_S = 0.3$ \\
            \hline
            ResNeXt-50~\cite{Qu:2019gqs} & 0.936 & 0.9837 & 302 $\pm$ 5 & 1147 $\pm$ 58 \\
            P-CNN~\cite{Qu:2019gqs} & 0.930 & 0.9803 & 201 $\pm$ 4 & 759 $\pm$ 24 \\
            PFN~\cite{Komiske:2018cqr} & - & 0.9819 & 247 $\pm$ 3 & 888 $\pm$ 17 \\
            ParticleNet~\cite{Qu:2019gqs} & 0.940 & 0.9858 & 397 $\pm$ 7 & 1615 $\pm$ 93\\
            JEDI-net~\cite{Moreno:2019bmu}  & 0.9300 & 0.9807 & - & 774.6 \\
            PCT~\cite{Mikuni:2021pou} & 0.940 & 0.9855 & 392 $\pm$ 11 & 1559 $\pm$ 98 \\
            LGN~\cite{Bogatskiy:2020tje} & 0.929 & 0.964 & - & 435 $\pm$ 95\\
            rPCN~\cite{Shimmin:2021pkm} & - & 0.9845 & 364 $\pm$ 9 & 1642 $\pm$ 93\\
            LorentzNet~\cite{Gong:2022lye} & 0.942 & 0.9868 & 498 $\pm$ 18 & 2195 $\pm$ 173 \\
            PELICAN~\cite{Bogatskiy:2022czk} & 0.9425 & 0.9869 & - & 2289 $\pm$ 204  \\
            ParT~\cite{Qu:2022mxj} & 0.940 & 0.9858 & 413 $\pm$ 16 &  1602 $\pm$ 81 \\
            ParT-f.t.~\cite{Qu:2022mxj} & \textbf{0.944} &  \textbf{0.9877} & \textbf{691 $\pm$ 15} & \textbf{2766 $\pm$ 130} \\
            Mixer~\cite{Hammad:2024cae} & - & 0.9859 & 416 & - \\
            \hline
            \textsc{PET} Classifier & 0.938 & 0.9848 & 340 $\pm$ 12 & 1318 $\pm$ 39\\
            \textsc{OmniLearn} & 0.942 & 0.9872 & 568 $\pm$ 9 & \textbf{2647 $\pm$ 192} \\
	\noalign{\smallskip}
	\end{tabular}
\end{table}

\begin{table}[th]
    \centering
    \caption{Comparison between the performance reported for different classification algorithms on the quark and gluon dataset. The uncertainty quoted corresponds to the standard deviation of nine trainings with different random weight initialization. If the uncertainty is not quoted then the variation is negligible compared to the expected value. Bold results represent the algorithm with highest performance. }
    \label{tab:results_qg}
	\begin{tabular}{lccccc}
    \noalign{\smallskip}\hline
          &  Acc &AUC & \multicolumn{2}{c}{1/$\epsilon_B$} \\
          \cline{4-5}
          & & & $\epsilon_S = 0.5$ & $\epsilon_S = 0.3$ \\
            \hline
            P-CNN~\cite{Qu:2019gqs} & 0.827 & 0.9002 & 34.7 & 91.0 \\
            PFN~\cite{Komiske:2018cqr} & - & 0.9005 & 34.7$\pm$0.4 & - \\
            ParticleNet~\cite{Qu:2019gqs} & 0.840 & 0.9116 & 39.8$\pm$0.2 & 98.6$\pm$1.3\\
            rPCN~\cite{Shimmin:2021pkm} & - & 0.9081 & 38.6 $\pm$ 0.5 & - \\
            ParT~\cite{Qu:2022mxj} & 0.840 & 0.9121 & 41.3 $\pm$ 0.3 & 101.2 $\pm$ 1.1 \\
            ParT-f.t.~\cite{Qu:2022mxj} & 0.843 &  0.9151 & 42.4 $\pm$ 0.2 & \textbf{107.9 $\pm$ 0.5} \\
            \hline
            \textsc{PET} classifier & 0.837 & 0.9110 & 39.92$\pm$0.1 & 104.9 $\pm$ 1.5\\
            \textsc{OmniLearn} & \textbf{0.844} & \textbf{0.9159} & \textbf{43.7$\pm$0.3} & \textbf{107.7 $\pm$ 1.5} \\
	\noalign{\smallskip}
	\end{tabular}
\end{table}

We first evaluate \textsc{OmniLearn} on two widely-used benchmark datasets for jet tagging: the top quark tagging~\cite{Kasieczka:2019dbj} and quark/gluon~\cite{Komiske:2018cqr} classification. In the top quark tagging dataset, events are simulated using \textsc{Pythia} 8 and \textsc{Delphes} with the ATLAS configuration.  The background consists of dijets produced via QCD and the signal consists of top quark pair production with all-hadronic decays.  The default energy flow algorithm in \textsc{Delphes} is used to create jet constituents, which are clustered using the anti-$k_T$ algorithm with $R=0.8$.  All jets in the range $550$~GeV~$< p_T < 650$~GeV and $|\eta| < 2$ are saved.  Note that while top quark and QCD categories are present in the JetClass dataset, the \textsc{Delphes} detector configuration and $p_T$ ranges are different. The quark/gluon dataset consists of stable particles clustered into jets, excluding neutrinos, using the anti-$k_{T}$ algorithm with radius R = 0.4. The quark-initiated sample (signal) is generated using a Z($\nu\nu$) + $(u,d,s)$ while the gluon-initiated data (background) are generated using Z($\nu\nu$) +$g$ processes. Both samples are generated using \textsc{Pythia8} without detector effects. Jets are required to have transverse momentum $\pt \in [500,550]$ GeV and rapidity $|y|<1.7$ for the reconstruction. For each dataset we evaluate the results of adapting \textsc{OmniLearn} on each dataset and compare with the training carried out with the \textsc{PET} classifier architecture from scratch. The results are compared with other models using the same datasets in Tables~\ref{tab:results_top} and \ref{tab:results_qg}. 

In both cases we observe the performance obtained by \textsc{OmniLearn} to be significantly better than other models trained from scratch, while matching and sometimes surpassing the performance observed by the fine-tuned version of the state-of-the-art model PartT~\cite{Qu:2022mxj}. In Figure.~\ref{fig:loss_top_qg}, we show the loss curve obtained during training in the validation set. After a single epoch, \textsc{OmniLearn} already reaches the performance from the \textsc{PET} classifier with convergence observed after only three epochs in both datasets, reducing the overall training time by a factor 3.

\section{Generalization across Detectors}
\label{sec:lhc}

\begin{figure}[th]
    \centering
        \includegraphics[width=.4\textwidth]{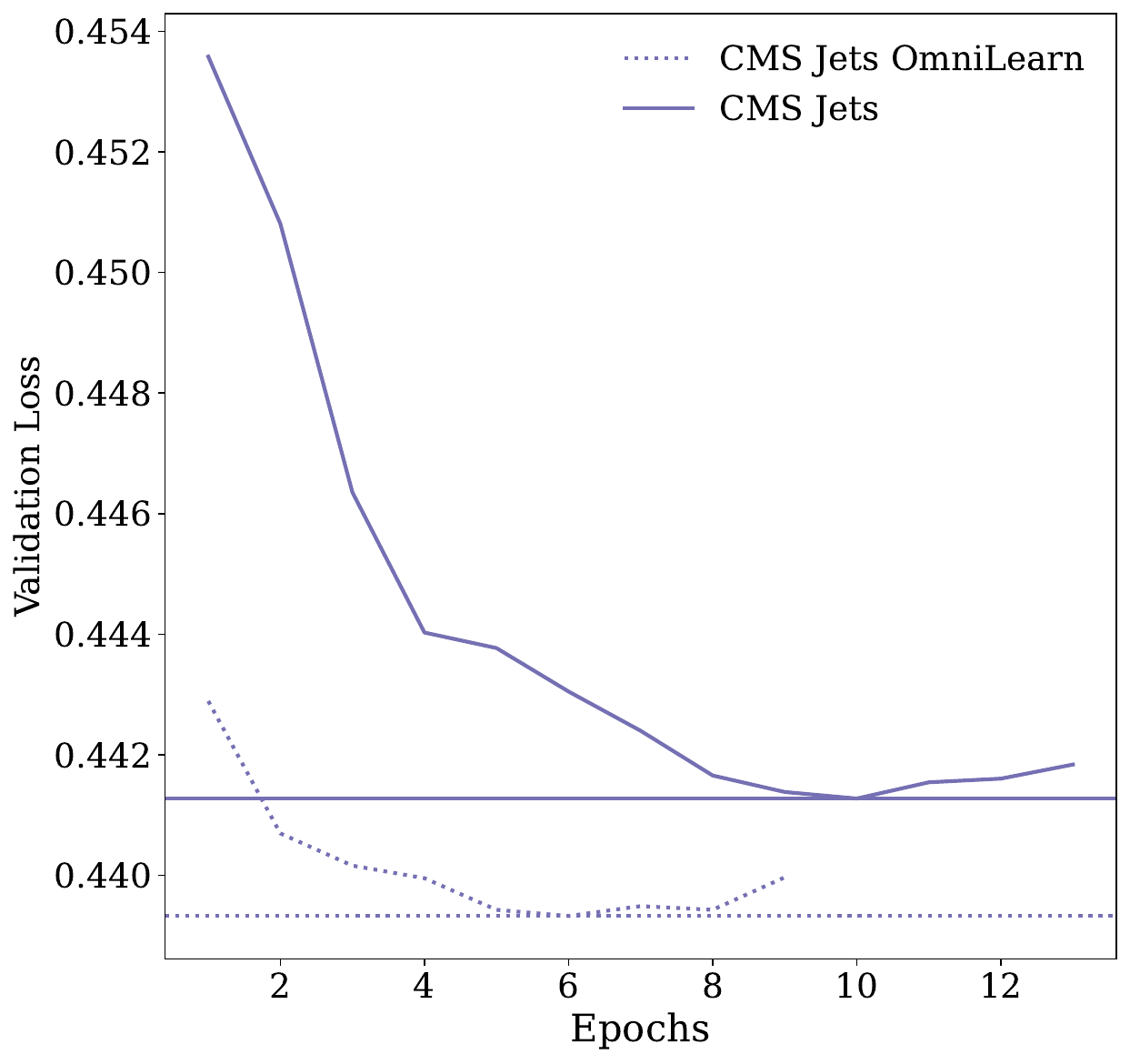}
    \caption{Validation loss curves obtained in the CMS Open Data quark gluon tagging dataset. The \textsc{OmniLearn} validation loss is compared with the \textsc{PET} classifier trained from scratch.}
    \label{fig:loss_cms}
\end{figure}

\begin{table}[th]
    \centering
    \caption{Comparison between the performance reported for different classification algorithms on the CMS  Open Data dataset to separate quark- from gluon-initiated jets. Bold results represent the algorithm with highest performance.}
    \label{tab:results_cms}
	\begin{tabular}{lccccc}
    \noalign{\smallskip}\hline
          &  AUC &Acc & \multicolumn{2}{c}{1/$\epsilon_B$} \\
          \cline{4-5}
          & & & $\epsilon_S = 0.5$ & $\epsilon_S = 0.8$ \\
            \hline
            \textsc{PET} classifier & 0.875  & 0.796 & 23.91 $\pm$ 0.07 & 4.770 $\pm$ 0.001\\
            \textsc{OmniLearn} & \textbf{0.877} & \textbf{0.797} & \textbf{24.36 $\pm$ 0.01} & \textbf{4.836 $\pm$ 0.004} \\
	\noalign{\smallskip}
	\end{tabular}
\end{table}

A more realistic scenario is to consider a complete simulation of the detector response. In this scenario, the generalization power from \textsc{OmniLearn}, trained on fast simulations, could be quickly adapted to studies using open dataset releases from the LHC Experiments. We investigate this scenario using the public CMS Open Data~\cite{CERN_Open_Data_Portal}. We use the simulations from the CMS Open Data release for 2011A run period~\cite{CMS_Jet_RunA_2011} processed by the MIT Open Data software~\cite{komiske_2019_3340205} to select simulated QCD jets in proton-proton collisions  at a center-of-mass energy of $\sqrt{s}$ = 7 TeV produced with \textsc{Pythia6}~\cite{Sjostrand:2006za} and Geant4~\cite{GEANT4:2002zbu} for detector effects. Particle-Flow objects~\cite{CMS:2017yfk} are used to define the objects that are clustered into jets using the anti-$k_{T}$ algorithm with radius parameter 0.5 with additional requirement to have $\pt >$ 375 GeV to achieve high trigger efficiency. The jet flavor is defined by the hard parton associated to the jet. We define quark-initiated jets as jets associated to uds partons and save these jets for the classification task against gluon-initiated jets. For simplicity, we fix the number of quark and gluon jets to be the same and around 20M events in total. We train \textsc{OmniLearn} using $70\%$ of the simulated jets, use $20\%$ for validation, and report the results compared to the \textsc{PET} classifier trained from scratch using a separate test set consisting of $10\%$ of the events in Table~\ref{tab:results_cms}. We observe the performance achieved by \textsc{OmniLearn} to be better in all metrics compared to a classifier trained from scratch. This observation is encouraging since \textsc{OmniLearn} shows that classification performance can be enhanced even when available datasets are of similar size as the ones used to train \textsc{OmniLearn}. Moreover, \textsc{OmniLearn} is able to converge 2 times faster and to a better minimum than training from scratch as shown by the progression of the validation loss in Figure~\ref{fig:loss_cms}.

\section{Generalization across Collision Systems}
\label{sec:h1}

\begin{figure}[ht]
    \centering
        \includegraphics[width=.4\textwidth]{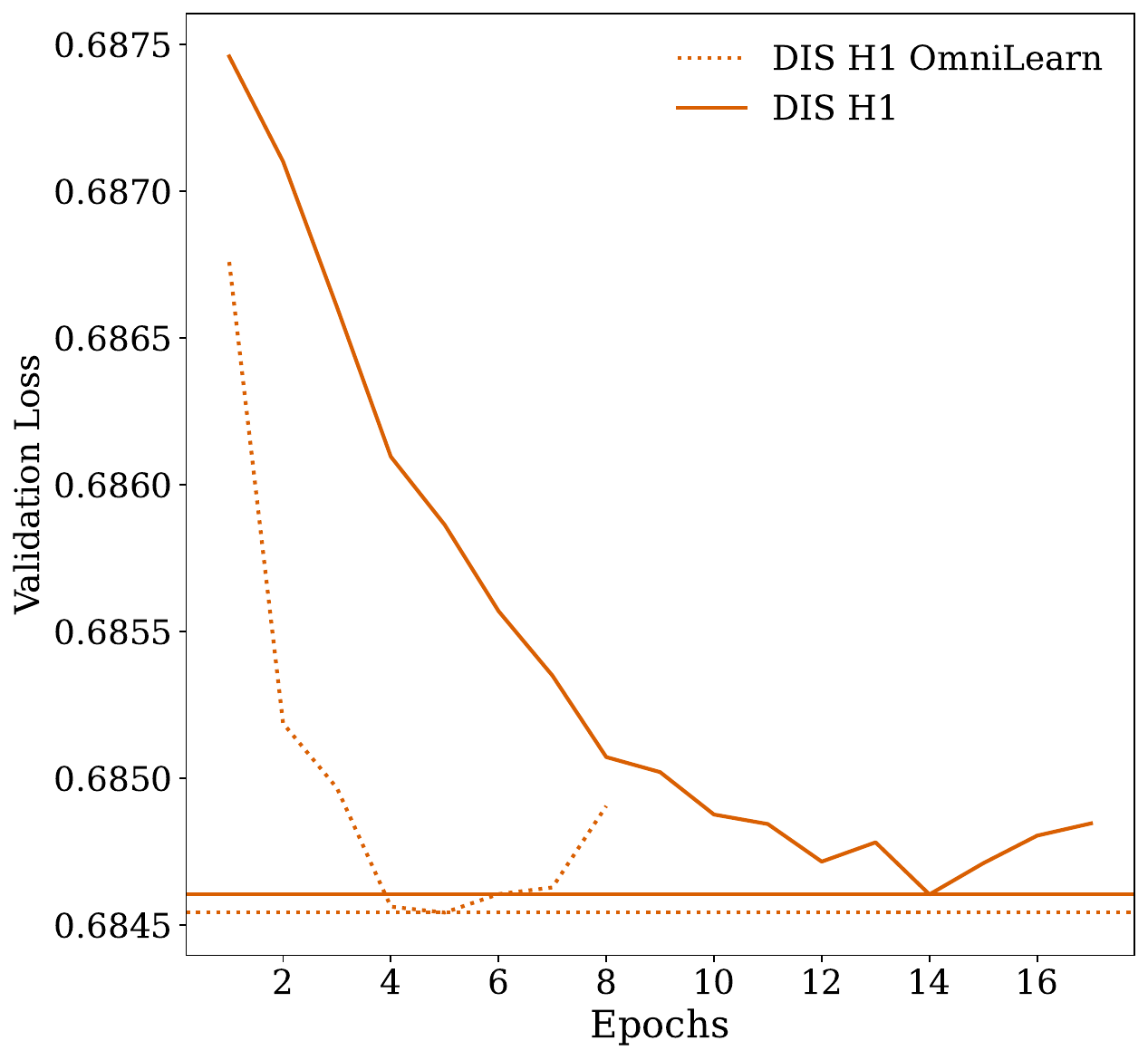}
    \caption{Validation loss curves obtained in the DIS dataset. The \textsc{OmniLearn} validation loss is compared with the \textsc{PET} classifier trained from scratch.}
    \label{fig:loss_h1}
\end{figure}

\begin{table}[ht]
    \centering
    \caption{Comparison between the performance reported for different classification algorithms on the DIS dataset. Both approaches show compatible performance within estimated uncertainties.}
    \label{tab:results_h1}
	\begin{tabular}{lccccc}
    \noalign{\smallskip}\hline
          &  AUC &Acc & \multicolumn{2}{c}{1/$\epsilon_B$} \\
         \cline{4-5}
         & & & $\epsilon_S = 0.1$ & $\epsilon_S = 0.5$ \\
            \hline
            \textsc{PET} classifier & 0.5691& 0.547& 17.73$\pm$0.04& 2.467$\pm$0.002\\
            \textsc{OmniLearn}  & 0.5695& 0.547& 17.78$\pm$0.06 & 2.470$\pm$0.003 \\
	\noalign{\smallskip}
	\end{tabular}
\end{table}

We also evaluate the generalization capability of \textsc{OmniLearn} for jets measured in different collision systems, covering completely different fiducial regions of the phase space compared to the LHC. We use simulations of neutral-current deep inelastic scattering (DIS) generated using the Rapgap 3.1~\cite{Jung:1993gf} generator for electron-proton collisions with electron and proton beam energies of 27.6 GeV and 920 GeV, respectively. The simulations are provided by the H1 Collaboration using the Heracles routines~\cite{Spiesberger:237380,Kwiatkowski:1990cx,Kwiatkowski:1990es} for QED radiation, CTEQ6L PDF set~\cite{Pumplin:2002vw}, and the Lund hadronization model~\cite{Andersson:1983ia}. The detector simulation is performed using the Geant3~\cite{Brun:1987ma} package. An energy-flow algorithm~\cite{energyflowthesis,energyflowthesis2,energyflowthesis3} is then used to reconstruct the particles clustered into jets using the $k_{T}$ algorithm with R=1.0. A second simulation using the Djangoh 1.4~\cite{Charchula:1994kf} generator is used during the classification. The task is to separate jets between the two different simulations. Notice that both simulations target the description of DIS events, hence their differences are more subtle than the previous classification tasks investigated thus far. This choice of classification problem is motivated by previous studies on the unfolding of jet substructure observables~\cite{H1:2023fzk}. One of the leading uncertainties, the closure test performed between two different simulations, is carried out using the same simulation routines and classification task used in this study, thus any improvements driven by \textsc{OmniLearn} could also lead to better unfolding algorithms to be developed for current and future unfolding analyses. A total of 2.5M events are generated for each simulation and used during the training. The performance obtained is listed in Table~\ref{tab:results_h1}.

We observe the same level of performance between \textsc{OmniLearn} and the \textsc{PET} classifier. Compared to the previous classification tasks, both accuracy and AUC values are much lower, evidencing the challenge of distinguishing jets from the two simulations apart. While the final performance is the same, we observe a much quicker convergence from \textsc{OmniLearn} as shown in the validation loss curve in Figure~\ref{fig:loss_h1}, resulting in a factor 3.5 faster training compared to starting from scratch. As described in~\cite{H1:2023fzk}, the full determination of the uncertainties require the training of thousands of classifiers, making any improvements in training speed an important asset for the applicability of the algorithm. These results also highlight the capability of \textsc{OmniLearn} to quickly adapt to new datasets, providing a strong asset for future experimental facilities such as the EIC.

\section{Conditional Generation}
\label{sec:jetnet}

\begin{figure*}[th]
    \centering
        \includegraphics[width=.4\textwidth]{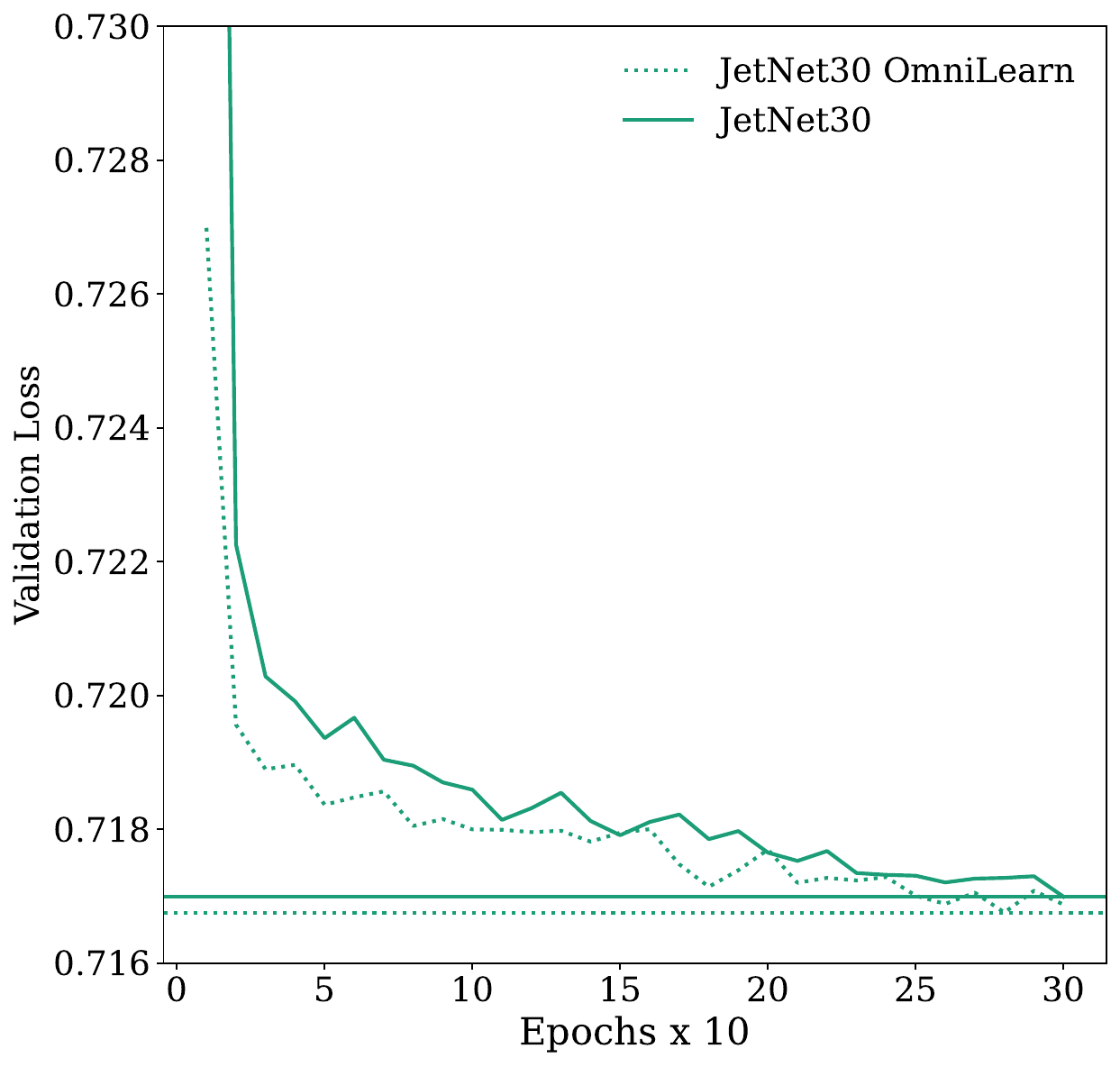}
        \includegraphics[width=.4\textwidth]{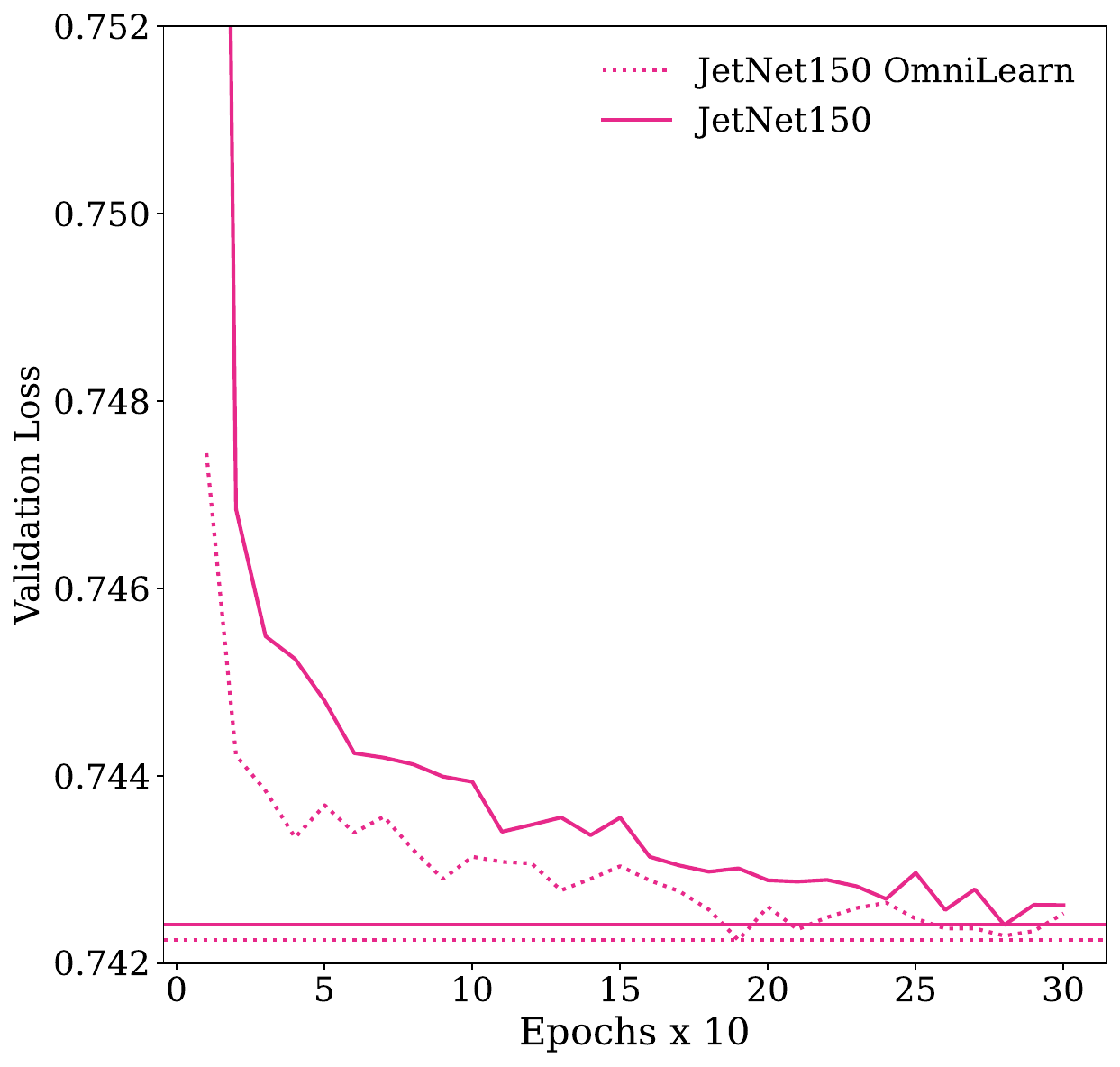}
    \caption{Validation loss curves obtained in the JetNet dataset with 30 particles (left) 150 particles. The \textsc{OmniLearn} validation loss is compared with the \textsc{PET} classifier trained from scratch.}
    \label{fig:loss_jetnet}
\end{figure*}

Next, we evaluate the generation quality obtained by \textsc{OmniLearn} using the JetNet~\cite{Kansal:2021cqp} datasets consisting of jets initiated by light-quarks, gluons, top quarks, $W$ and $Z$ bosons. The jets are generated with transverse momenta $\pt$ around 1~TeV and are clustered using the anti-$k_t$ algorithm with a radius parameter of 0.4. Each jet has a maximum number of particles stored fixed to 30~\cite{kansal_raghav_2022_6975118} or 150~\cite{kansal_raghav_2022_6975117}. For each jet, the four-momentum information $(\pt_\mathrm{jet},\eta_\mathrm{jet},\phi_\mathrm{jet}, m_\mathrm{jet})$ is provided, as well as the particle multiplicity. We adopt the two-model strategy presented in~\cite{Mikuni:2023dvk}, training a model that only learns the kinematic information of the jets and using that information as conditional information for the diffusion model trained using particles as inputs. Note that in this case, \textsc{OmniLearn} is used only to learn the particle information while the overall jet kinematic information model is always trained from scratch. The performance of the generation is evaluated using multiple physics-based metrics proposed in~\cite{Kansal:2021cqp} and listed in Tables~\ref{tab:jetnet30} and~\ref{tab:jetnet150}.
We also provide the results of \textsc{OmniLearn} and the \textsc{PET} generator in the ideal case, where the jet information, used to condition the particle generation model, is taken directly from the validation set of the JetNet dataset, effectively separating the impact of the jet generation, that does not benefit from \textsc{OmniLearn}, from the particle generation process.

\begin{table*}[th]
\centering
\caption{Comparison of the results obtained between different generative models in the task of particle property generation in the dataset consisting of 30 particles. Lower is better for all metrics except Cov. FPND metrics are not available for W and Z bosons, hence omitted.}
\label{tab:jetnet30}
%\begin{adjustbox}{width=1\textwidth}
\begin{tabular}{ | c | c | c | c | c | c | c | c | c | c | c | c | c |}
\hline
Jet class & Model & W$^\mathrm{PM}_1$ ($\times 10^{-3}$) & W$^\mathrm{P}_1$ ($\times 10^{-3}$) & W$^\mathrm{PEFP}_1$ ($\times 10^{-5}$) & FPND & Cov$\uparrow$ & MMD \\
\hline

          & FPCD~\cite{Mikuni:2023dvk} & \textbf{0.36 $\pm$ 0.08 }& \textbf{0.34 $\pm$ 0.09} & 0.47 $\pm$ 0.13& 0.07 & 0.55 & 0.03 \\
    Gluon & FPCD 1~\cite{Mikuni:2023dvk} & 0.65 $\pm$ 0.11 & \textbf{0.34 $\pm$ 0.06} & 0.60 $\pm$ 0.09 & 0.11 & 0.55 & 0.03  \\
          & \textsc{MP-GAN}~\cite{Kansal:2021cqp} & 0.69 $\pm$ 0.07 & 1.8 $\pm$ 0.2 &0.9 $\pm$ 0.6 &0.20 &0.54 &0.037 \\
          & \textsc{EPiC-GAN}~\cite{Buhmann:2023pmh} & \textbf{0.3 $\pm$ 0.1} & 1.6 $\pm$ 0.2& 0.4 $\pm$ 0.2 & 1.01 $\pm$ 0.07& -& -\\
          & \textsc{PET} generator & 0.42 $\pm$ 0.10 & \textbf{0.36 $\pm$ 0.08 }& \textbf{0.35 $\pm$ 0.08} & 0.04 & 0.55 & 0.03  \\
          & \textsc{PET} generator (Ideal) & \textbf{0.36 $\pm$ 0.08} & \textbf{0.34 $\pm$ 0.09} & 0.47 $\pm$ 0.13& 0.07 & 0.55 & 0.03 \\
          & \textsc{OmniLearn} & \textbf{0.38 $\pm$ 0.08} & \textbf{0.33 $\pm$ 0.07} & \textbf{0.33 $\pm$ 0.09} & \textbf{0.02} & 0.55 & 0.03  \\
          &\textsc{OmniLearn} (Ideal) & \textbf{0.33 $\pm$ 0.06} & \textbf{0.29 $\pm$ 0.08} & \textbf{0.30 $\pm$ 0.07} & \textbf{0.02 }& 0.55 & 0.03\\
\hline
          & FPCD~\cite{Mikuni:2023dvk} &  0.52 $\pm$ 0.07 & \textbf{0.27 $\pm$ 0.06} & 0.38 $\pm$ 0.11 & 0.08 & 0.49 & 0.02  \\
Light Quark& FPCD 1~\cite{Mikuni:2023dvk} & 0.59 $\pm$ 0.08 & 0.36 $\pm$ 0.08 & 0.50 $\pm$ 0.08 & 0.09 & 0.48 & 0.02 \\
          & \textsc{MP-GAN}~\cite{Kansal:2021cqp} & 0.6 $\pm$ 0.2 & 4.9 $\pm$ 0.5 & 0.7 $\pm$ 0.4 & 0.35 &  0.50 &  0.026\\
          & \textsc{EPiC-GAN}~\cite{Buhmann:2023pmh} & 0.5 $\pm$ 0.1 & 4.0 $\pm$ 0.4 & 0.8 $\pm$ 0.4 & 0.43 $\pm$ 0.03 & -& -\\
          & \textsc{PET} generator & 0.39 $\pm$ 0.12 & 0.35 $\pm$ 0.06 & \textbf{0.24 $\pm$ 0.10} & 0.03 & \textbf{0.54} & 0.02 \\
          & \textsc{PET} generator (Ideal) & 0.31 $\pm$ 0.08 & 0.38 $\pm$ 0.10 & \textbf{0.23 $\pm$ 0.07} & 0.03 & 0.53 & 0.02\\
          & \textsc{OmniLearn} & \textbf{0.24 $\pm$ 0.03} & \textbf{0.32 $\pm$ 0.07} & \textbf{0.24 $\pm$ 0.08} & 0.02 & \textbf{0.54} & 0.02 \\
          &\textsc{OmniLearn} (Ideal) & 0.31 $\pm$ 0.08 & \textbf{0.30 $\pm$ 0.09} & \textbf{0.26 $\pm$ 0.08} & \textbf{0.01} & \textbf{0.54} & 0.02 \\
\hline
          & FPCD~\cite{Mikuni:2023dvk} & 0.51 $\pm$ 0.07 & 0.41 $\pm$ 0.12 & 1.25 $\pm$ 0.19 & 0.17 & 0.58 & 0.05 \\
Top Quark & FPCD 1~\cite{Mikuni:2023dvk} & 1.22 $\pm$ 0.09 &\textbf{ }0.46 $\pm$ 0.10 & 2.66 $\pm$ 0.26 & 0.56 & 0.57 & 0.05 \\
          & \textsc{MP-GAN}~\cite{Kansal:2021cqp} & 0.6 $\pm$ 0.2 & 2.3 $\pm$ 0.3 & 2 $\pm$ 1 & 0.37 & 0.57 & 0.071 \\
          & \textsc{EPiC-GAN}~\cite{Buhmann:2023pmh} & 0.5 $\pm$ 0.1 & 2.1 $\pm$  0.1 & 1.7 $\pm$ 0.3 & 0.31 $\pm$ 0.037 & -& -\\
          & \textsc{PET} generator & 0.44 $\pm$ 0.03 & \textbf{0.29 $\pm$ 0.07} & \textbf{1.09 $\pm$ 0.23} & 0.07 & 0.58 & 0.05 \\
          & \textsc{PET} generator (Ideal) & \textbf{0.41 $\pm$ 0.07} & \textbf{0.34 $\pm$ 0.08} & \textbf{1.22 $\pm$ 0.23} & 0.07 & 0.58 & 0.05\\
          & \textsc{OmniLearn} & 0.43 $\pm$ 0.06 & \textbf{0.30 $\pm$ 0.07} & 1.31 $\pm$ 0.18 & 0.04 & 0.58 & 0.05 \\
          &\textsc{OmniLearn} (Ideal) & \textbf{0.36 $\pm$ 0.05} & 0.41 $\pm$ 0.08 & \textbf{1.02 $\pm$ 0.20} & \textbf{0.03} & 0.58 & 0.05 \\
\hline
          & FPCD~\cite{Mikuni:2023dvk} & 0.26 $\pm$ 0.03 & 0.39 $\pm$ 0.08 & 0.15 $\pm$ 0.02 & - & 0.56 & 0.02 \\
W Boson   
          & FPCD 1~\cite{Mikuni:2023dvk} & 0.94 $\pm$ 0.06 & 0.42 $\pm$ 0.09 & 0.35 $\pm$ 0.03 & - & 0.56 & 0.02 \\
          & \textsc{PET} generator & \textbf{0.17 $\pm$ 0.04} & \textbf{0.26 $\pm$ 0.05} & \textbf{0.11 $\pm$ 0.02} & - & 0.56 & 0.02 \\
          & \textsc{PET} generator (Ideal) & \textbf{0.15 $\pm$ 0.02} & \textbf{0.31 $\pm$ 0.07} & \textbf{0.12 $\pm$ 0.03} & - & \textbf{0.57} & 0.02 \\
          & \textsc{OmniLearn} & \textbf{0.19 $\pm$ 0.03} & \textbf{0.27 $\pm$ 0.07 }& \textbf{0.10 $\pm$ 0.02} & - & \textbf{0.57} & 0.02 \\
          &\textsc{OmniLearn} (Ideal) & \textbf{0.16 $\pm$ 0.06} & \textbf{0.28 $\pm$ 0.04 }& \textbf{0.10 $\pm$ 0.02} & - & \textbf{0.57} & 0.02  \\

\hline
          &  FPCD~\cite{Mikuni:2023dvk} & \textbf{0.21 $\pm$ 0.04} & 0.40 $\pm$ 0.13 & 0.18 $\pm$ 0.03 & - &  0.56 & 0.02 \\
Z Boson   
          & FPCD 1~\cite{Mikuni:2023dvk} & 0.99 $\pm$ 0.05 & 0.35 $\pm$ 0.06 & 0.49 $\pm$ 0.03 &- & 0.56 & 0.02  \\
          & \textsc{PET} generator & \textbf{0.22 $\pm$ 0.04} & \textbf{0.32 $\pm$ 0.07} & 0.20 $\pm$ 0.04 & - &  \textbf{0.57} & 0.02 \\
          & \textsc{PET} generator (Ideal) & \textbf{0.18 $\pm$ 0.10} & \textbf{0.30 $\pm$ 0.08} & \textbf{0.14 $\pm$ 0.02} & - & 0.56 & 0.02 \\
          & \textsc{OmniLearn} & \textbf{0.19 $\pm$ 0.07 }& \textbf{0.32 $\pm$ 0.09} & \textbf{0.12 $\pm$ 0.03} & - &  \textbf{0.57} & 0.02 \\
          &\textsc{OmniLearn} (Ideal) & \textbf{0.22 $\pm$ 0.05} & \textbf{0.27 $\pm$ 0.06} & \textbf{0.13 $\pm$ 0.02} &  - & \textbf{0.57} & 0.02 \\
\hline

\end{tabular}
%\end{adjustbox}
\end{table*}

\begin{table*}[th]
\centering
\caption{Comparison of the results obtained between different generative models in the task of particle property generation in the dataset consisting of 150 particles. Baseline FPCD~\cite{Mikuni:2023dvk} uses 512 time steps during sampling. Distilled models are listed alongside number of time steps used. Lower is better for all metrics except Cov. }
\label{tab:jetnet150}
%\begin{adjustbox}{width=1\textwidth}
\begin{tabular}{ | c | c | c | c | c | c | c | c | c | c | c | c | c |}
\hline
Jet class & Model & W$^\mathrm{PM}_1$ ($\times 10^{-3}$) & W$^\mathrm{P}_1$ ($\times 10^{-3}$) & W$^\mathrm{PEFP}_1$ ($\times 10^{-5}$) & Cov$\uparrow$ & MMD \\
\hline
          & FPCD~\cite{Mikuni:2023dvk} & 0.44 $\pm$ 0.11 & \textbf{0.28 $\pm$ 0.05} & 0.91 $\pm$ 0.16 & \textbf{0.56} & 0.03 \\
    Gluon & FPCD 1~\cite{Mikuni:2023dvk} & 0.65 $\pm$ 0.12 & 0.58 $\pm$ 0.03 & 1.49 $\pm$ 0.34 & 0.55 & 0.03 \\
          & \textsc{EPiC-GAN}~\cite{Buhmann:2023pmh} & 0.4 $\pm$ 0.1 & 3.2 $\pm$ 0.2 & 1.1 $\pm$ 0.7 & -& -\\
          & \textsc{PET} generator & \textbf{0.32 $\pm$ 0.09} & 0.34 $\pm$ 0.07 & 1.19 $\pm$ 0.26 & \textbf{0.56} & \textbf{0.02} \\
          & \textsc{PET} generator (Ideal) & \textbf{0.32 $\pm$ 0.18} & 0.29 $\pm$ 0.07 & 0.95 $\pm$ 0.41 & \textbf{0.56} & \textbf{0.02 }\\
          & \textsc{OmniLearn} & 0.47 $\pm$ 0.18 & 0.31 $\pm$ 0.11 & 1.05 $\pm$ 0.23 & \textbf{0.56} & \textbf{0.02} \\
          &\textsc{OmniLearn} (Ideal) & \textbf{0.32 $\pm$ 0.09} & \textbf{0.22 $\pm$ 0.06} & \textbf{0.65 $\pm$ 0.20} & 0.55 & \textbf{0.02} \\
\hline
          & FPCD~\cite{Mikuni:2023dvk} &  0.46 $\pm$ 0.05 & \textbf{0.24 $\pm$ 0.02} & \textbf{0.43 $\pm$ 0.09} & 0.54 & 0.02 \\
Light Quark& FPCD 1~\cite{Mikuni:2023dvk} & \textbf{0.39 $\pm$ 0.04} & 0.61 $\pm$ 0.03 & 0.57 $\pm$ 0.10 & 0.54 & 0.02 \\
          & \textsc{EPiC-GAN}~\cite{Buhmann:2023pmh} & \textbf{0.4 $\pm$ 0.1} & 3.9 $\pm$ 0.3 & 0.7 $\pm$ 0.4 & -& -\\
          & \textsc{PET} generator & \textbf{0.41 $\pm$ 0.04} & 0.34 $\pm$ 0.08 & 0.74 $\pm$ 0.18 & \textbf{0.55} & 0.02 \\
          & \textsc{PET} generator (Ideal) &\textbf{ 0.34 $\pm$ 0.09} & 0.34 $\pm$ 0.12 & \textbf{0.50 $\pm$ 0.17} & \textbf{0.55} & 0.02  \\
          & \textsc{OmniLearn} & 0.46 $\pm$ 0.13 & 0.39 $\pm$ 0.11 & 0.54 $\pm$ 0.14 & 0.53 & 0.02 \\
          &\textsc{OmniLearn} (Ideal) & \textbf{0.34 $\pm$ 0.15} & 0.41 $\pm$ 0.11 & \textbf{0.41 $\pm$ 0.12} & 0.54 & 0.02 \\
\hline
          & FPCD~\cite{Mikuni:2023dvk} & 0.40 $\pm$ 0.07 & \textbf{0.30 $\pm$ 0.03} & 2.23 $\pm$ 0.16 & \textbf{0.58} & 0.05 \\
Top Quark & FPCD 1~\cite{Mikuni:2023dvk} & 0.85 $\pm$ 0.09 & 0.87 $\pm$ 0.03 & 3.82 $\pm$ 0.24 & \textbf{0.58} & 0.05 \\
          & \textsc{EPiC-GAN}~\cite{Buhmann:2023pmh} & 0.6 $\pm$ 0.1 & 3.7 $\pm$ 0.3 & 2.8 $\pm$ 0.7 & -& -\\
          & \textsc{PET} generator & 0.40 $\pm$ 0.08 & \textbf{0.28 $\pm$ 0.08} & 1.81 $\pm$ 0.33 & 0.57 & \textbf{0.04} \\
          & \textsc{PET} generator (Ideal) & \textbf{0.29 $\pm$ 0.07} & 0.36 $\pm$ 0.05 & \textbf{1.27 $\pm$ 0.30 }& 0.57 & \textbf{0.04} \\
          & \textsc{OmniLearn} & 0.38 $\pm$ 0.05 & \textbf{0.30 $\pm$ 0.07} & 1.84 $\pm$ 0.30 & 0.57 & \textbf{0.04} \\
          &\textsc{OmniLearn} (Ideal) & \textbf{0.30 $\pm$ 0.07} & \textbf{0.28 $\pm$ 0.07} & \textbf{1.16 $\pm$ 0.39} & 0.57 & \textbf{0.04} \\
          
\hline
          & FPCD~\cite{Mikuni:2023dvk} & 0.29 $\pm$ 0.02 &\textbf{ }\textbf{0.23 $\pm$ 0.02} & 0.22 $\pm$ 0.04 & 0.55 & 0.02 \\
W Boson   & FPCD 1~\cite{Mikuni:2023dvk} & 0.93 $\pm$ 0.04 & 0.67 $\pm$ 0.01 & 0.37 $\pm$ 0.03 & \textbf{0.56} & 0.02 \\
          & \textsc{PET} generator & \textbf{0.15 $\pm$ 0.02} & 0.27 $\pm$ 0.07 & \textbf{0.12 $\pm$ 0.03} & 0.55 & 0.02 \\
          & \textsc{PET} generator (Ideal) & \textbf{0.12 $\pm$ 0.03} & \textbf{0.24 $\pm$ 0.06} & \textbf{0.13 $\pm$ 0.03} & 0.55 & 0.02 \\
          & \textsc{OmniLearn} & 0.18 $\pm$ 0.01 & 0.27 $\pm$ 0.05 & \textbf{0.14 $\pm$ 0.04} & \textbf{0.56} & 0.02 \\
          &\textsc{OmniLearn} (Ideal) & \textbf{0.13 $\pm$ 0.04} & \textbf{0.26 $\pm$ 0.04} & \textbf{0.11 $\pm$ 0.03} & 0.55 & 0.02 \\

\hline
          & FPCD~\cite{Mikuni:2023dvk} & 0.28 $\pm$ 0.05 & \textbf{0.22 $\pm$ 0.03} & 0.23 $\pm$ 0.03 & 0.55 & 0.02 \\
Z Boson   & FPCD 1~\cite{Mikuni:2023dvk} & 1.04 $\pm$ 0.08 & 0.69 $\pm$ 0.02 & 0.62 $\pm$ 0.06 & \textbf{0.57} & 0.02 \\
          & \textsc{PET} generator & 0.24 $\pm$ 0.06 & 0.35 $\pm$ 0.06 & 0.20 $\pm$ 0.04 & 0.55 & 0.02 \\
          & \textsc{PET} generator (Ideal) & \textbf{0.13 $\pm$ 0.02} & \textbf{0.22 $\pm$ 0.06} & \textbf{0.16 $\pm$ 0.04} & 0.55 & 0.02 \\
          & \textsc{OmniLearn} & 0.19 $\pm$ 0.05 & 0.38 $\pm$ 0.11 & 0.19 $\pm$ 0.03 & 0.56 & 0.02 \\
          &\textsc{OmniLearn} (Ideal) & \textbf{0.12 $\pm$ 0.03} & 0.28 $\pm$ 0.09 & \textbf{0.14 $\pm$ 0.04} & 0.55 & 0.02 \\
\hline

\end{tabular}
%\end{adjustbox}
\end{table*}

In all metrics investigated in this study, \textsc{OmniLearn} shows similar or improved performance compared to previous models for both datasets consisting of 30 and 150 particles. We also notice that differences between the idealized version of \textsc{OmniLearn} and \textsc{PET} generator are more significant than at full generation level where the jet generation quality also affects the overall model performance. Additionally, statistical uncertainties determined from the bootstrapping method are often above the 20$\%$ level for some metrics, requiring access to larger evaluation sample sizes to provide a more precise comparison between models. In Figure~\ref{fig:loss_jetnet} we show the validation loss curve for the \textsc{PET} generator and \textsc{OmniLearn} training in the JetNet dataset. Similarly to previous results, the \textsc{OmniLearn} training starts from a lower value of the loss function and is able to converge quicker, requiring roughly 20$\%$ and 30$\%$ fewer training epochs than a model trained from scratch. The generation time is dominated by the number of time-steps used in the diffusion process. Both \textsc{PET} and \textsc{OmniLearn} have the same generation time and are roughly a factor 2 faster compared to the FPCD~\cite{Mikuni:2023dvk} model as using half as many steps as FPCD already saturates the performance within uncertainties. Notice that similarly to FPCD, we can accelerate the inference time through the use of techniques like distillation~\cite{salimans2022progressive}.

\section{Likelihood Ratio Estimation}
\label{sec:omnifold}

\begin{figure*}[th]
    \centering
        \includegraphics[width=.3\textwidth]{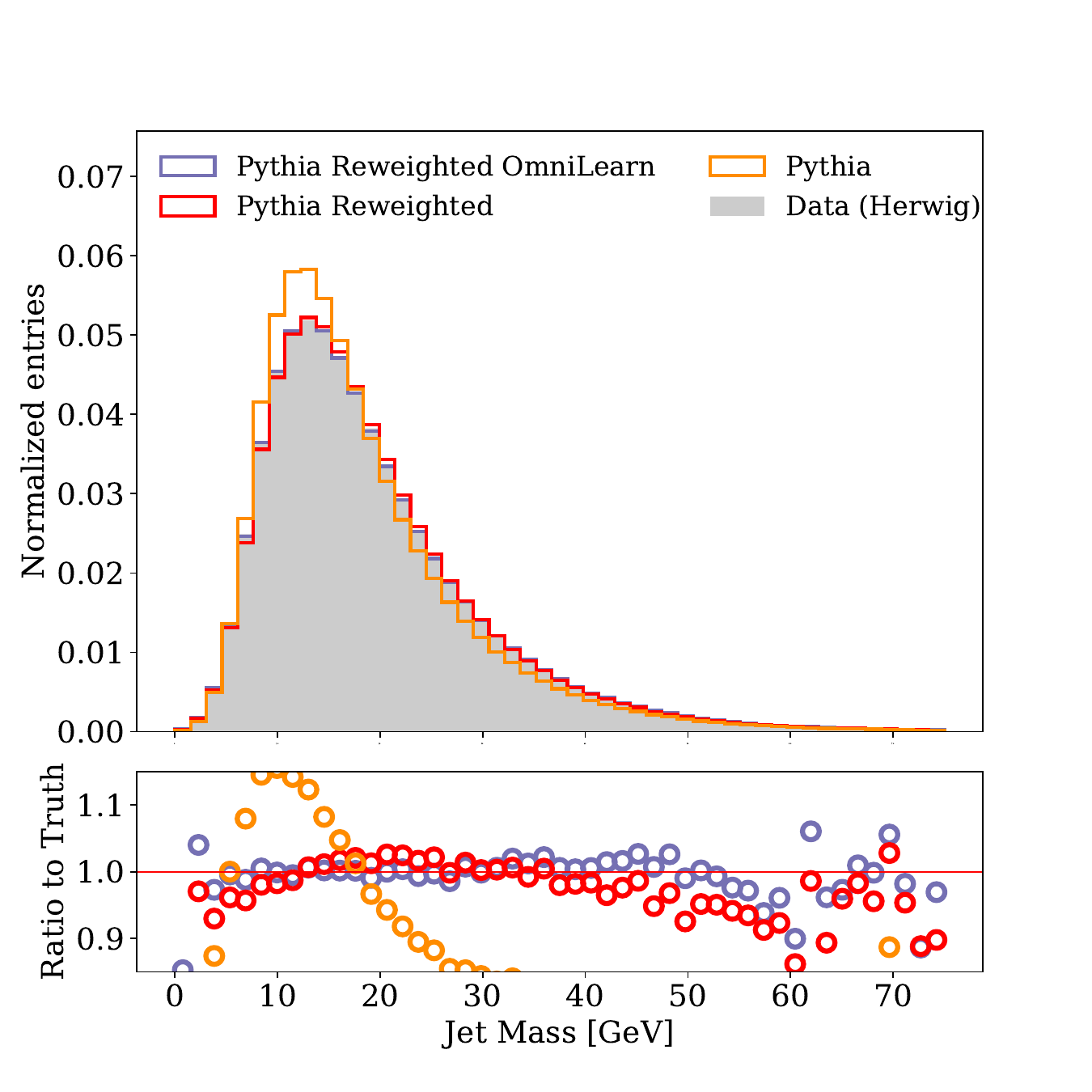}
        \includegraphics[width=.3\textwidth]{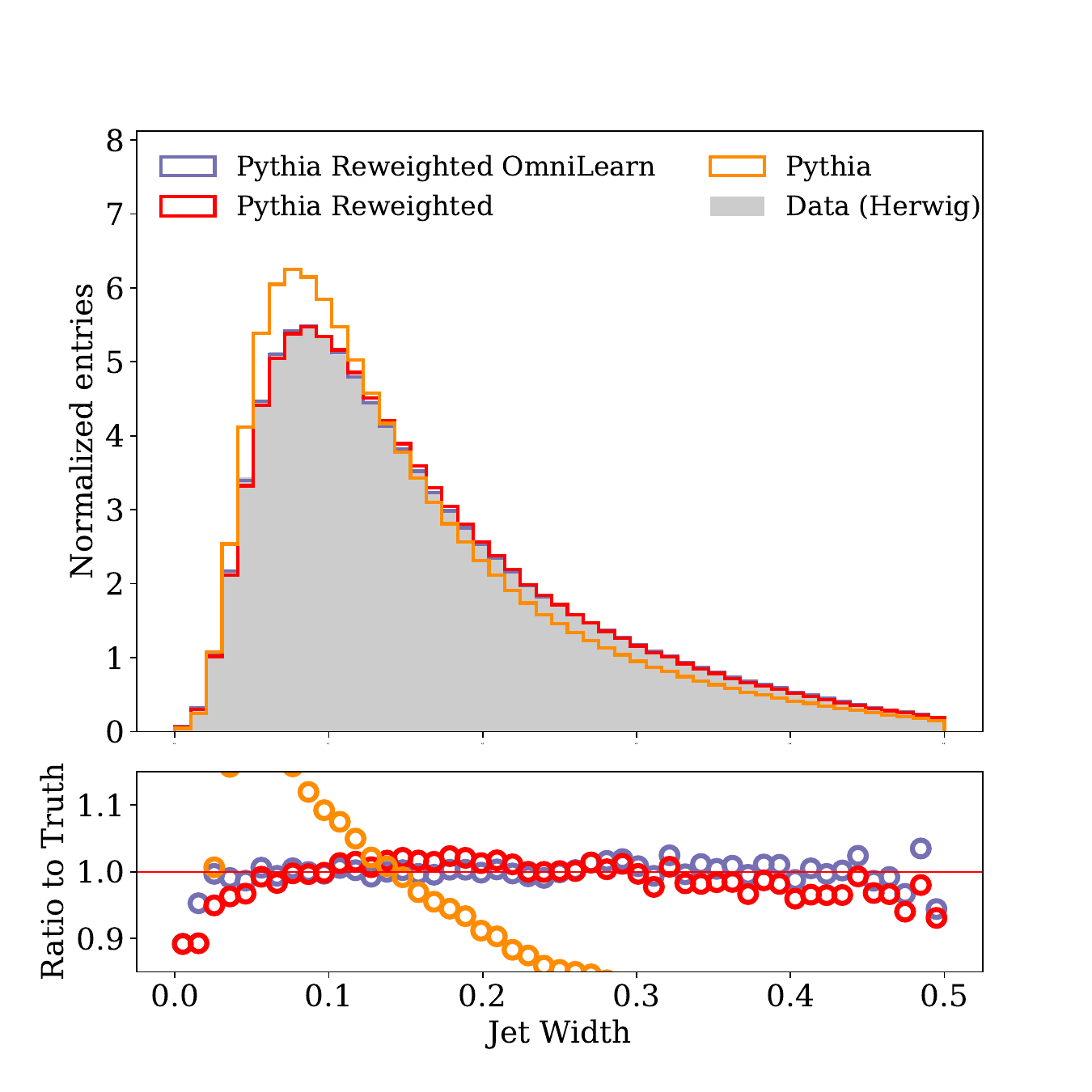}
        \includegraphics[width=.3\textwidth]{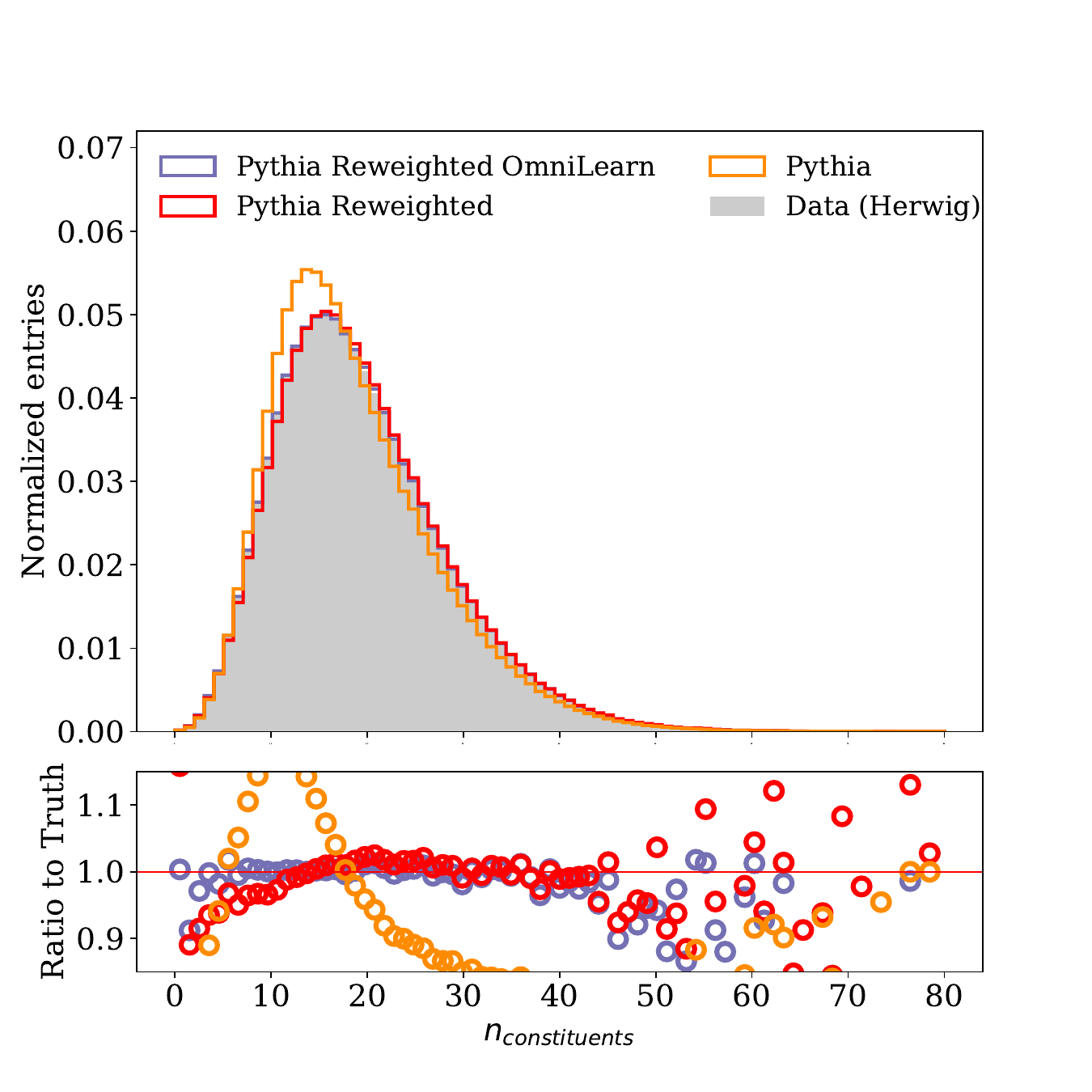}
        \includegraphics[width=.3\textwidth]{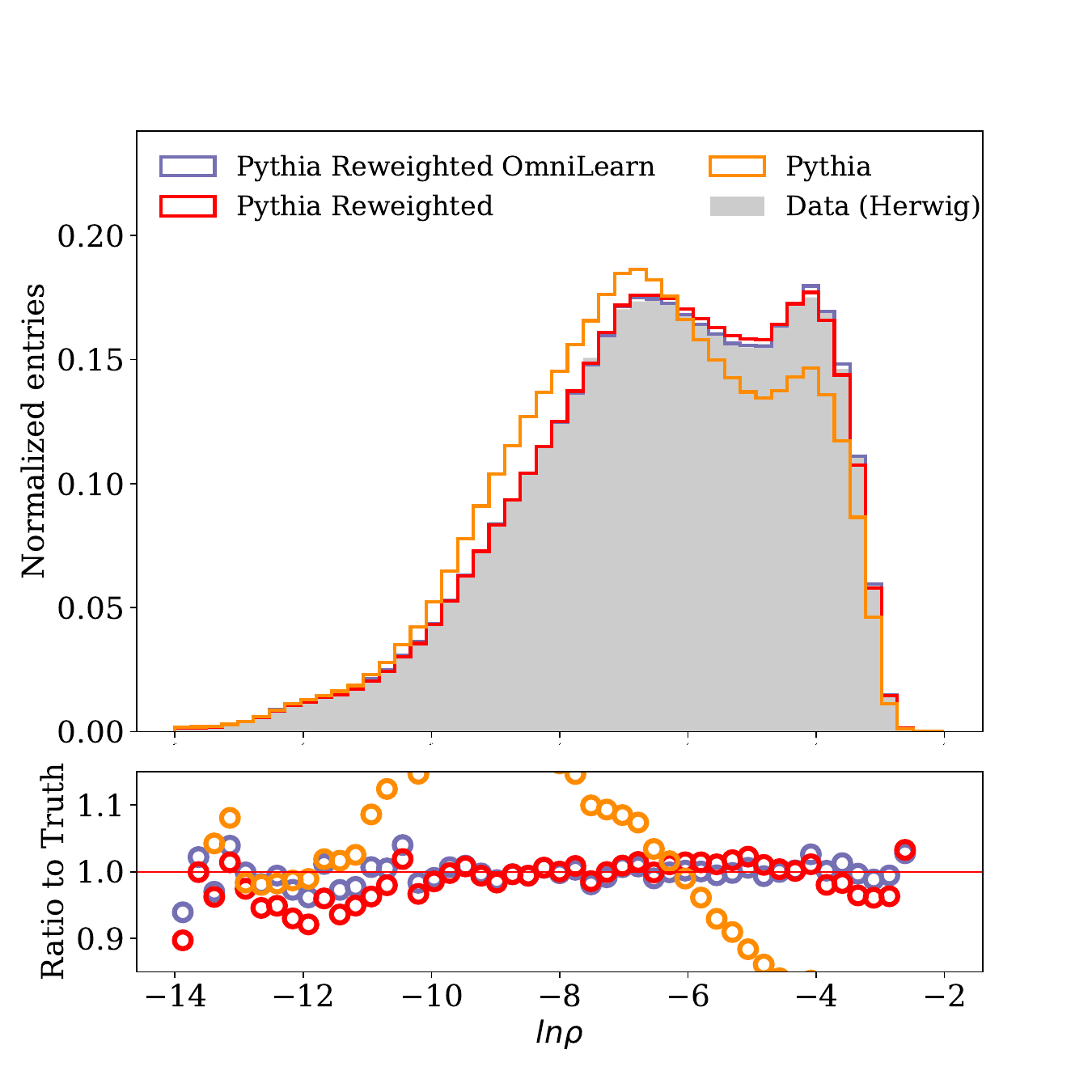}
        \includegraphics[width=.3\textwidth]{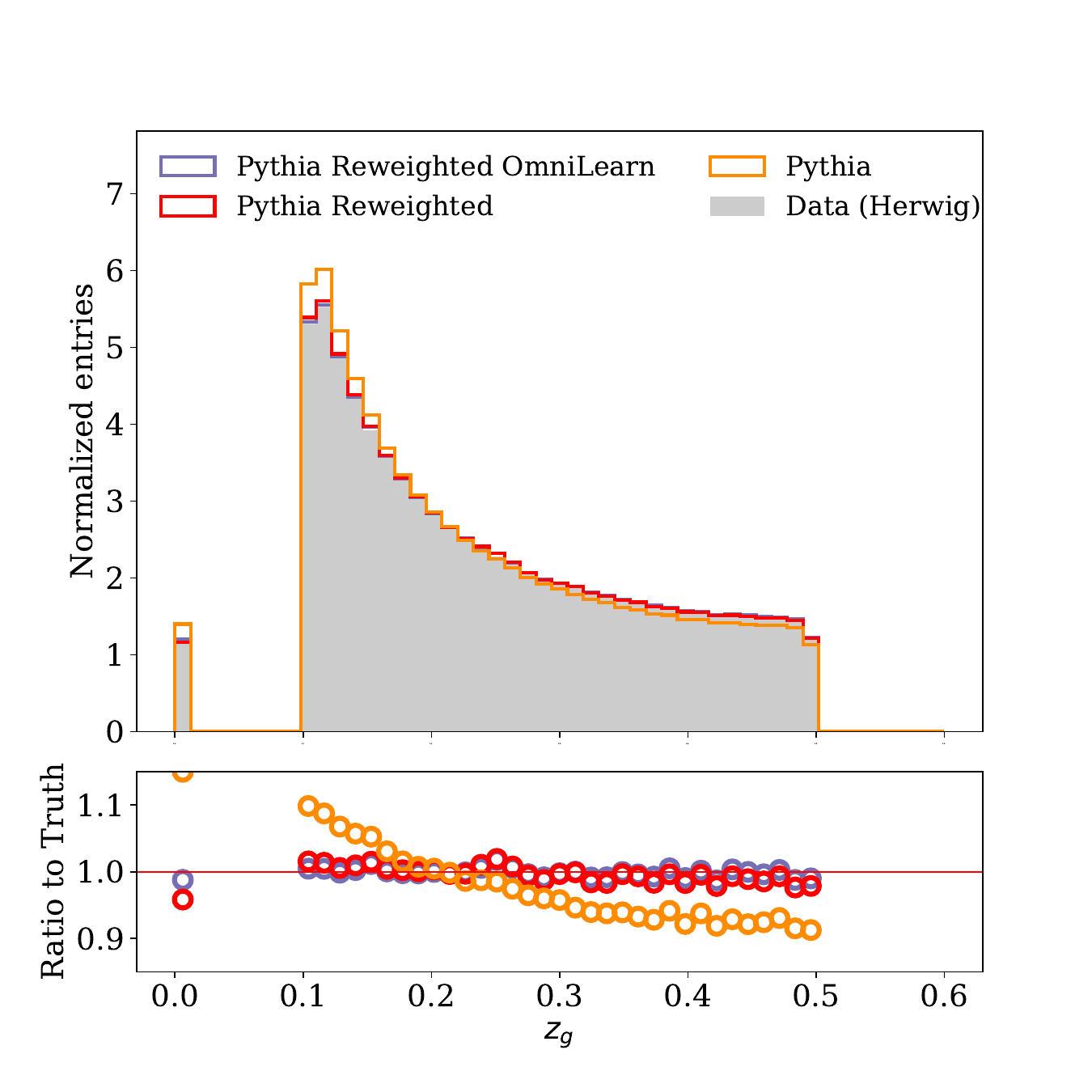}
        \includegraphics[width=.3\textwidth]{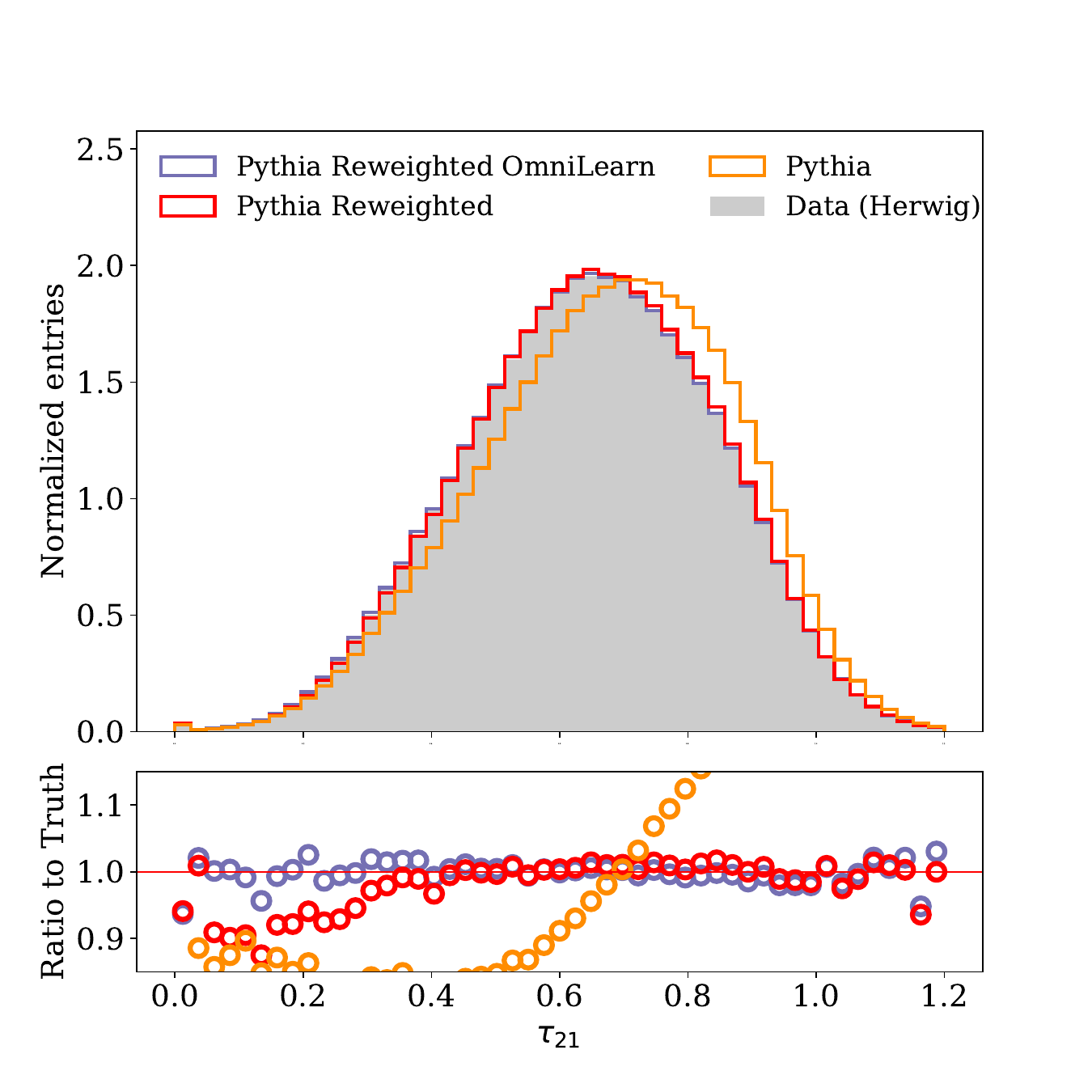}
    \caption{Reweighted distributions for six different physics observables obtained from the \textsc{OmniLearn} and \textsc{PET} classifier.}
    \label{fig:reweighted}
\end{figure*}

Reweighting distributions is a common task in collider physics, often used to correct physics observables based on measured data. Different methods using non-deep learning algorithms were proposed to reweight a discrete number of observables~\cite{LHCb:2017zzt,Martschei:2012pr, Rogozhnikov:2016bdp,ATLAS:2018tti}. More recently, parameterized classifiers~\cite{Cranmer:2015bka,Andreassen:2019nnm} were shown to achieve precise event reweighting, even in high dimensional spaces. The classifier prediction is used to determine the reweighting function by approximating the likelihood ratio between samples. We examine the potential of \textsc{OmniLearn} applied to high-dimensional reweighting considering all available features and the same dataset introduced in Ref.~\cite{Andreassen:2019cjw}, available on Zenodo~\cite{andreassen_anders_2019_3548091}. The dataset consists of proton-proton collisions producing a $Z$ boson, generated at a center-of-mass energy of $\sqrt{s}=14$ TeV. A sample used as the `data' representative is simulated using particle collisions with the default tune of Herwig 7.1.5~\cite{Bahr:2008pv,Bellm:2015jjp,Bellm:2017bvx}. A second dataset, representative of the `simulation' we want to correct, is simulated using \textsc{Pythia8} with Tune 21~\cite{ATL-PHYS-PUB-2014-021}. Detector distortions are simulated with \textsc{Delphes} and the CMS tune that uses a particle flow reconstruction. Jets are clustered using all particle flow objects at detector level and all stable non-neutrino truth particles at particle level.  They are defined by the anti-$k_T$ algorithm with radius parameter $R=0.4$ as implemented in FastJet~3.3.2~\cite{Cacciari:2011ma}. The $Z$ bosons are required to have $\pt>200$ GeV in order to mitigate acceptance effects. While all clustered particles are used during the training, we report the performance obtained by different algorithms using the same set of observables reported in the original \textsc{OmniFold} publication. 

We investigate the capability of \textsc{OmniLearn} to determine the full phase space reweighting function by fine-tuning the classifier. We also compare with the results obtained from a classifier trained from scratch (\textsc{PET} classifier). The results of the reweighted distributions at reconstruction level are shown in Figure~\ref{fig:reweighted}. From a visual inspection we see that both \textsc{OmniLearn} and the \textsc{PET} classifier are able to correctly reweight the distributions of all high-level observables we investigated, however \textsc{OmniLearn} shows a better agreement with the `data'. We quantify the improvement brought by \textsc{OmniLearn} by calculating the triangular discriminator~\cite{850703,Gras:2017jty,Bright-Thonney:2018mxq} for each observable with results reported in Table~\ref{tab:omnifold_reco}. In all cases, the values obtained by \textsc{OmniLearn} are significantly better than the baseline training.

\begin{table}[ht]
    \centering
    \caption{Comparison of the triangular discriminator between different algorithms for reweighting. Uncertainties from \textsc{PET} and \textsc{OmniLearn} are taken from 100 histogram variations within the statistical uncertainty of the prediction. Quantities in bold represent the method with best performance.}
    \label{tab:omnifold_reco}
    \begin{tabular}{lcc}
        & \textsc{PET} classifier & \textsc{OmniLearn} \\
        \hline
        Jet mass & 0.13$\pm$0.03 & \textbf{0.027$\pm$0.008} \\
        N & 0.13$\pm$0.03 & \textbf{0.05$\pm$0.02} \\
        Jet Width & 0.09$\pm$0.02 & \textbf{0.02$\pm$0.01} \\
        $\log\rho$ & 0.08$\pm$0.02 & \textbf{0.03$\pm$0.01} \\
        $\tau_{21}$ & 0.08$\pm$0.03 & \textbf{0.02$\pm$0.01} \\
        $z_g$ & 0.04$\pm$0.01 & \textbf{0.001$\pm$0.004} \\
        \end{tabular}
    \end{table}

\section{Weak Supervision and Resonant Anomaly Detection}
\label{sec:lhco}

The search for new particle interactions is a challenging task at collider experiments. Expected to be rare, signs for new physics might be subtle, affecting only very specific observables. While traditional approaches rely on theory predictions to narrow down possible new physics scenarios, the current lack of new physics discoveries at the LHC motivates new strategies to search for new phenomena. Anomaly detection brings a change in this paradigm~\cite{Kasieczka:2021xcg,Aarrestad:2021oeb,Karagiorgi:2021ngt}. Unexpected data structures can be automatically identified by algorithms as possible hints for new physics. One well-studied approach designed for resonant anomaly detection is based on Classification Without Labels (CWoLa) framework~\cite{Metodiev:2017vrx,Collins:2018epr,Collins:2019jip}, where weakly-supervised learning enables training directly on (unlabeled) data. In the CWoLa approach, samples with mixed fractions of a possible signal and background are used to train a classifier whose goal is to identify the origin of the sample. In the best case, we can imagine a sample consisting of only the  background process, possibly as part of a background simulation or derived from a control region, that is then used to train the classifier against data possibly containing new particle interactions in addition to the background process. We examine the benefits of using \textsc{OmniLearn} during the classification process and evaluate the performance using the R\&D dataset from the LHC Olympics data challenge~\cite{LHCOlympics,Kasieczka:2021xcg}.  The background consists of dijet final states from QCD production while the signal is a resonant boson production $A \to B (\to q q') C(\to q q') $ with masses $m_{A}, m_B, m_C$ = 3.5, 0.5, 0.1~TeV, respectively.  Signal and background vents are generated with \textsc{Pythia8} interfaced with \textsc{Delphes3.4.1} for detector simulation. Jets are defined using the anti-$k_T$ algorithm as implemented in \textsc{FastJet} with $R=1$. We focus on the two leading jets in transverse momentum space and require the leading jet to have $\pt > 1.2$~TeV. After selection, we save all particles associated to the two most energetic jets, resulting in a maximum particle multiplicity of 279 particles per jet. 

\begin{figure}[ht]
    \centering
        \includegraphics[width=.4\textwidth]{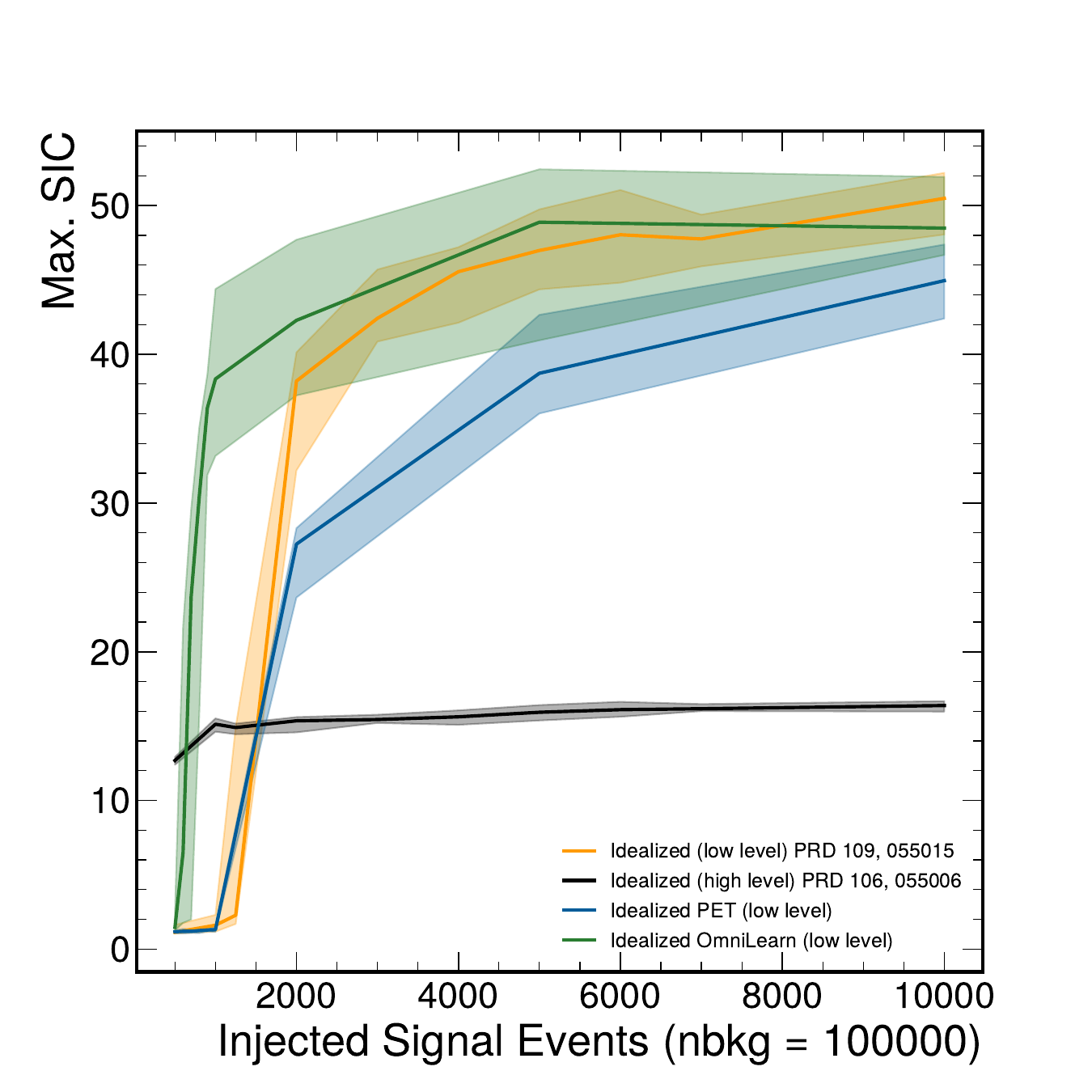}
    \caption{Maximum values of the SIC curve evaluated over different values of injected signal. \textsc{OmniLearn} and \textsc{PET} classifier results are compared with other algorithms used for the same task.}
    \label{fig:max_sic_ideal}
\end{figure}

During weakly-supervised training, we follow previous studies~\cite{Nachman:2020lpy,Andreassen:2020nkr,Hallin:2021wme,Benkendorfer:2020gek,Kasieczka:2021tew,Hallin:2022eoq,Sengupta:2023xqy,Raine:2022hht,Golling:2023yjq,Bickendorf:2023nej,Das:2023bcj,Finke:2023ltw,Freytsis:2023cjr,Buhmann:2023acn} and define the signal region of interest for events with dijet mass $3300~\text{GeV} < m_{jj} < 3700~\text{GeV}$. In the region of interest, we call `data' the combination of 100k background events with varying amounts of signal events. The background-only distribution consists of 350k independently simulated background events. Differently to previous applications of \textsc{OmniLearn} described in this paper, here, we need to include the information of both jets in the classifier. We modify the \textsc{PET} classifier to accommodate the changes in the dataset while also preserving the permutation equivariance of the complete network. This is achieved by first passing the particles present in each jet through the \textsc{PET} body independently, such that particles belonging to different jets do not interact with each other. The outputs of the \textsc{PET} body are then shifted and scaled by the outputs of the jet embedding block that takes as input the kinematic information of each jet. This strategy allows us to maintain the permutation equivariance of the model while giving jet specific information to each particle. The shifted and scaled particles are then passed to the classifier head, reshaped as if all particles belonged to the same jet. Since the shift and scaling operations are jet-dependent, the reshaping operation allows all particles to be conditionally mapped to the same space without loss of information. The classifier head is unchanged, with a class token used to summarize the information of all particles before the classification output. We use the output of the classifier as the anomaly score to determine the sensitivity to this specific new physics scenario. We quantify the performance based on the maximum value of the significance improvement characteristic curve (SIC) defined as the signal efficiency divided by the square root of the background efficiency versus the signal efficiency. The SIC represents a multiplicative factor by which the initial significance of a signal present in the data would increase when a particular threshold of the classifier output is chosen.  Maximum SIC values above unity indicate value added. We show the results in Figure~\ref{fig:max_sic_ideal} and compare the results obtained by \textsc{OmniLearn} and \textsc{PET} classifier with the results reported in~\cite{Buhmann:2023acn}. Since the reference background process is statistically identical to the background presented in the `data' construction, we call this scenario idealized\footnote{The authors of Ref.~\cite{Das:2023bcj} show that it is possible to achieve even better performance if the functional form is known - it would be interesting to see such a strategy combined with \textsc{OmniLearn} in the future.}.  

\textsc{OmniLearn} shows non-negligible signal sensitivity for signals injections above 600, corresponding to an initial significance $S/\sqrt{B}\sim 2$, representing a large increase in sensitivity compared to  previous results where signal sensitivity was only achieved for signal injections above 1500 ($S/\sqrt{B}\sim 5$). Compared to previous results, we also observe the performance of \textsc{PET} classifier to be similar at lower signal injections to the results reported in~\cite{Buhmann:2023acn} and performing worse at higher signal injections. The reason for this difference is due to the limited amount of data in the signal region (around 100k). Even with the larger generated background of 350k events, the dataset size of this application is at least 2 times smaller than all previous datasets investigated so far. In the low data regimes, data efficient models are often observed to perform better than large transformer models, a limitation that is mitigated by \textsc{OmniLearn}.

\section{Conclusion and Outlook}
\label{sec:conclusions}

In this paper, we have introduced a foundation model for jet physics called \textsc{OmniLearn}.  This machine learning approach is a neural network capable of advancing a wide variety of research areas within jet physics.  We have shown that even though this model was trained on one specific dataset, it accelerates and/or improves the accuracy of other classification tasks and generation quality evaluated over eight additional datasets across initial states, final states, and simulation levels as well as of tasks in jet generation, likelihood estimation, and anomaly detection.  While our neural network has two million trainable parameters, it is much smaller than many other foundation models (i.e. it is not `large' in the sense of an LLM).  We view this as a strength for usability and we hypothesize that this is possible because our model had a training task close to the application tasks. 
Moreover, \textsc{OmniLearn} has the potential to overcome many of the current challenges in collider physics associated to training taggers on computationally expensive simulations, high-dimensional event unfolding, and full-event anomaly detection. These key applications are explored in~\cite{prl}.

Our methodology and trained model are publicly available.  While we have focused on jet physics, it would be exciting to see \textsc{OmniLearn} applied in other, related areas with similar challenges and thus also similar potential rewards.

\section*{Code Availability}

The code for this paper can be found at \url{https://github.com/ViniciusMikuni/OmniLearn}.

\section*{Acknowledgments}
We thank F. Dreyer, Ming Fong, R. Grabarczyk, K. Greif, P.F. Monni, and D. Whiteson for interesting discussions related to transfer learning.  We also thank J. Birk,  A. Hallin , and G. Kasieczka for comments on the manuscript
Additionally, we thank our colleagues from the H1 Collaboration for allowing us to use the simulated MC event samples. We also thank DESY-IT and the MPI f\"ur Physik for providing computing infrastructure and supporting the data preservation project of the HERA experiments.
VM, and BN are supported by the U.S. Department of Energy (DOE), Office of Science under contract DE-AC02-05CH11231.  This research used resources of the National Energy Research Scientific Computing Center, a DOE Office of Science User Facility supported by the Office of Science of the U.S. Department of Energy under Contract No. DE-AC02-05CH11231 using NERSC awards HEP-ERCAP0021099 and HEP-ERCAP0028249.

\appendix

\section{Input Variables}
\label{app:inputs}
The input features used to train \textsc{OmniLearn} are described in Table~\ref{tab:features}.

\begin{table}[th]
    \centering
    \caption{Input features used during the \textsc{OmniLearn} training. Binary flags consisting of charge and PID information are included in the `is *' observables.}
    \label{tab:features}
	\begin{tabular}{lc}
    \hline\noalign{\smallskip} 
             Object & Observables  \\
            \hline
            & $\Delta\eta$  \\  
            & $\Delta\phi$ \\  
            & $\log\pt$\\  
            & $\log(1 - \frac{\pt}{\pt(\mathrm{jet})})$\\  
            & $\log\mathrm{E}$\\          
            & $\log(1 - \frac{\mathrm{E}}{\mathrm{E}(\mathrm{jet})})$ \\  
           Particles  & $\Delta\mathrm{R}$\\  
            & charge\\  
            & is electron\\
            & is muon\\
            & is photon\\
            & is charged hadron\\
            & is neutral hadron\\
            \hline
            &  $\pt$\\
            Jets &  $\eta$\\
             &  mass \\
            &  particle multiplicity\\

    \noalign{\smallskip}\hline  
	\end{tabular}
\end{table}

\bibliography{HEPML,other}
\bibliographystyle{apsrev4-1}

\clearpage

\end{document}